\documentclass[showpacs,preprintnumbers,amssymb,twocolum,aps,reprint,superscriptaddress,showkeys,amsmath,floatfix]{revtex4-2}

\usepackage{graphicx}
\usepackage{dcolumn}
\usepackage[dvipsnames]{xcolor}
\usepackage{bm}
\usepackage{amsmath}
\usepackage{amssymb}
\usepackage{epsfig}
\usepackage{amsfonts}
\usepackage{lineno,hyperref}
\usepackage{array}
\usepackage{float}
\usepackage{microtype}
\usepackage{multirow}
\usepackage{adjustbox}
\usepackage[english]{babel}
\usepackage{epstopdf}
\usepackage{blindtext}
\usepackage{booktabs}
\usepackage{subcaption}
\usepackage[a4paper, total={6.8in, 10in}]{geometry}
\usepackage{appendix}

\def \be{\begin{equation}}
\def \ee{\end{equation}}
\def \ben{\begin{eqnarray}}
\def \een{\end{eqnarray}}

\def \La{\mathcal{L}}

\def \n{\nonumber\\}

\begin{document}

\title{Gravitational Wave Propagation in K-essence Cosmology: Theory and Observational Constraints}

\author{Sougata Bhunia}
\email{sougatabhunia066@gmail.com}
\affiliation{Department of Physics, University of Calcutta, 92, A.P.C. Road, Kolkata-700009, India}

\author{Goutam Manna$^a$}
\email{goutammanna.pkc@gmail.com \\$^a$Corresponding author}
\affiliation{Department of Physics, Prabhat Kumar College, Contai, Purba Medinipur 721404, India} 
\affiliation{Institute of Astronomy, Space and Earth Science, Kolkata 700054, India}

\author{Eduardo Guendelman}
\email{guendel@bgu.ac.il}
\affiliation{Department of Physics, Ben-Gurion University of the Negev, Beer-Sheva, Israel}
\affiliation{Frankfurt Institute for Advanced Studies (FIAS), Ruth-Moufang-Strasse 1, 60438 Frankfurt am Main, Germany.}
\affiliation{Bahamas Advanced Study Institute and Conferences, 4A Ocean Heights, Hill View Circle, Stella Maris, Long Island, The Bahamas.}

\author{Debashis Gangopadhyay}
\email{debashis.g@snuniv.ac.in}
\affiliation{Department of Physics, School of Natural Sciences, Sister Nivedita University, DG 1/2, Action Area 1, Newtown, Kolkata 700156, India.}

\author{Ramón Herrera}
\email{ramon.herrera@pucv.cl}
\affiliation{Instituto de F\'{\i}sica, Pontificia Universidad Cat\'{o}lica de Valpara\'{\i}so, Avenida Brasil 2950, Casilla 4059, Valpara\'{\i}so, Chile.}

\author{Abhijit Bhattacharyya}
\email{abhattacharyyacu@gmail.com}
\affiliation{Department of Physics, University of Calcutta, 92, A.P.C. Road, Kolkata-700009, India}

\date{\today}

\begin{abstract}
Gravitational waves (GWs) provide a powerful, theory-independent probe of the dynamical structure of spacetime and the cosmological background. We investigate the propagation of linearized tensor modes in k-essence cosmology, where the dark sector is described by a non-canonical scalar field. Working in the high-frequency (short-wavelength) limit on a Friedmann-Lema\^{\i}tre-Robertson-Walker background and restricting to the transverse-traceless sector, we obtain a modified evolution equation in which the GW speed remains strictly luminal (compatible with GW170817), while the background field $\bar{\phi}$ generates a time-dependent effective mass term $m_{\rm eff}$. This background-induced mass alters the dispersion relation without introducing extra propagating degrees of freedom, producing an accumulated, frequency-dependent phase shift over cosmological distances. We show that $m_{\rm eff}$ is uniquely determined by background cosmological parameters and can be expressed as a redshift-dependent function $m_{\rm eff}(z)$, providing a direct connection between GW observables and scalar-field dynamics. The GW luminosity distance stays identical to the electromagnetic one, preserving standard-siren consistency. We confront the model with a joint Bayesian analysis combining cosmic chronometers, BAO, Pantheon+SH0ES, and standard-siren measurements from GWTC-2.1/3/4, and assess performance using $\chi^2$, $\chi^2_{\rm red}$, AIC, BIC, and DIC. The reconstruction is consistent with current data, reproduces the late-time expansion history, and provides data-driven constraints on the redshift evolution of $m_{\rm eff}(z)$.

\end{abstract}

%%%%%%%%%%%%
%\pacs{04.60.−m, 04.70.Dy, 04.62.+v}
%%%%%%%%%%%%%
\maketitle

\section{Introduction} 
The theoretical prediction of gravitational waves (GWs) dates back to the formulation of General Relativity (GR), where Einstein showed that the linearized field equations admit propagating perturbations of the spacetime metric \cite{Albert}. Later work placed gravitational radiation on firm physical ground by clarifying its generation and energy transport in curved spacetime \cite{Bondi, Sachs}. Indirect evidence was provided by the observed orbital decay of the binary pulsar PSR~B1913+16, in remarkable agreement with the GR prediction \cite{Taylor, Hulse}. A century after their prediction, the first direct detection---the binary black-hole merger GW150914---opened the era of gravitational-wave astronomy \cite{Abbott3}. Since then, successive GW catalogs \cite{GWTC1, GWTC2, GWTC2.1, GWTC3, GWTC4} have established gravitational physics as an observational science and provided unique access to the dynamical sector of spacetime. Beyond their astrophysical impact, GWs also offer powerful cosmological probes. In particular, standard sirens \cite{Shutz1, Holz} enable direct distance measurements that are independent of the electromagnetic distance ladder, providing a complementary route to constrain the expansion history of the universe and the properties of its dominant components.

Within GR, gravitational waves can be described as massless tensor perturbations propagating on a background spacetime: they satisfy a dispersionless wave equation and travel at the speed of light \cite{Maggiore, Eanna}. In a cosmological setting based on the Friedmann--Lema\^itre--Robertson--Walker (FLRW) metric, the expansion of the universe primarily affects GWs through an adiabatic damping of their amplitude as they travel over cosmological distances. This picture has been tested stringently by the multi-messenger observation of the binary neutron-star merger GW170817 \cite{Abbott} and its electromagnetic counterpart GRB~170817A \cite{Abbott1}, which constrains any deviation between the propagation speed of GWs and light to be extremely small. However, while such measurements severely limit modifications to the speed of gravity, they do not exclude subtler propagation effects induced by the cosmological background, in particular scenarios in which the dispersion properties of the waves are modified without changing their (low-frequency) propagation speed \cite{Belgacem1, Nishizawa}.

A broader and more systematic framework for scalar--tensor theories is provided by Horndeski gravity, the most general scalar--tensor theory with second-order field equations and free from Ostrogradsky instabilities \cite{Horndeski, Deffayet, Kobayashi}. In this class of theories, derivative couplings between the scalar field and curvature can modify both the background cosmological evolution and the propagation of gravitational waves. Extensive work has explored these effects, showing that departures from GR may appear in the GW propagation speed, the effective amplitude damping, and the dispersion relation \cite{Belgacem1, Nishizawa, Lombriser}. The tight bound on the GW speed from GW170817 \cite{Abbott}, however, excludes large regions of the Horndeski parameter space, leaving viable only subclasses that predict (effectively) luminal propagation \cite{Creminelli, Ezquiaga, Baker}. Within the surviving sector, k-essence (a subclass of the Horndeski gravity) provides a well-motivated limit in which the Lagrangian depends only on the scalar field and its kinetic term, with no explicit non-minimal coupling to curvature \cite{Picon2, Picon, Garriga}. In this limit, modifications to GW propagation arise through the time-dependent cosmological background sourced by the scalar field.

The presence of such a non-trivial background raises a natural question: how does the dark sector influence the propagation of gravitational waves? Since GWs travel across cosmological distances, even small departures from standard propagation can accumulate into observable effects. In particular, modifications to the effective dispersion relation can induce frequency-dependent phase shifts in the waveform, providing a sensitive probe of the underlying cosmological medium \cite{Will, Mirshekari}.\\

In this work, we study the linearized propagation of gravitational waves on a cosmological background sourced by a k-essence scalar field. A central theoretical challenge is the consistent definition of perturbations on a curved, time-dependent spacetime, where the split between background and fluctuations is not unique. We address this issue by working in \emph{the short-wavelength (high-frequency) approximation}, in which GWs are rapidly varying perturbations with wavelength much smaller than the characteristic curvature scale of the background \cite{Isaacson, Isaacson1}. Within this framework, we derive the linearized Einstein equations in the presence of k-essence, treating the metric and scalar field as independent dynamical degrees of freedom. By adopting unitary gauge (suppressing intrinsic scalar perturbations), we isolate the impact of the k-essence background on the tensor sector. We show that, although the GW propagation speed remains strictly luminal, the combined effects of cosmological expansion and the non-canonical scalar dynamics generate an effective mass-like term in the GW equation.

We consider models described by a non-canonical Lagrangian $\mathcal{L}(X,\phi)$. For transverse--traceless tensor modes on an FLRW background, we derive the modified evolution equation and show that the effective mass alters the dispersion relation without introducing additional propagating degrees of freedom. This modification accumulates into a frequency-dependent phase shift in GW signals over cosmological distances, providing an observational signature absent in standard GR and qualitatively distinct from scenarios that modify the propagation speed or the amplitude damping \cite{Belgacem1, Ezquiaga}. Importantly, the effective mass can be expressed in terms of cosmological quantities such as the Hubble rate and the scalar-field equation of state, thereby linking GW observables directly to the physics of dark energy.

Overall, our results underscore the potential of gravitational waves as precision probes of the cosmological background, since tensor perturbations can carry subtle imprints of the dark sector. As GW measurements gain sensitivity and extend to higher redshift, such effects provide a promising route to constrain the microphysics of dark energy and to test departures from the standard cosmological model. In contrast to general Horndeski or EFT-of-dark-energy frameworks, where the tensor kinetic sector can be modified and the GW speed altered, the present minimally coupled k-essence scenario preserves strictly luminal propagation while inducing a purely background-generated dispersive correction through an effective tensor potential. This work isolates that mechanism and relates it directly to cosmological observables.\\

The paper is organized as follows. In Section~\ref{Sec.II}, we present the theoretical framework for gravitational-wave propagation in curved spacetime. In Section~\ref{Sec.III}, we introduce the k-essence scalar-field model, derive the background cosmological equations, and review linear perturbations on an FLRW background. In Section~\ref{Sec.IV}, we obtain the modified gravitational-wave evolution equation in the presence of the k-essence field, highlighting the emergence of an effective mass term and its implications for the dispersion relation. In Section~\ref{Sec.V}, we describe our methodology for confronting the model with observational data. In Section~\ref{Sec:VI}, we fit the model to several datasets---\textit{Cosmic Chronometers} (CC), \textit{Pantheon+SHOES}, \textit{Baryon Acoustic Oscillations} (BAO), and \textit{Gravitational-Wave Transient Catalogs} (GWTC; GWTC-2.1, GWTC-3, and GWTC-4)---and discuss the results. Section~\ref{Sec.VII} concludes with a summary of our findings.

\section{Review of Linearized Gravity theory for curved background spacetime} \label{Sec.II}
Our analysis begins by formulating the equation for metric perturbation in a general curved and dynamical background spacetime, the solutions of which ultimately yield the gravitational wave profile for the given geometry. That means in our consideration, the background stress-energy tensor is non-zero. So, the perturbed metric can be written as \cite{Maggiore, Maggiore1, Wald, Carroll, Schutz, Blau}
\ben
g_{\mu\nu}(x) = \bar{g}_{\mu\nu}(x) + h_{\mu\nu}(x),
\qquad ||h_{\mu\nu}|| \ll 1 .
\label{1}
\een
where $h_{\mu\nu}(x)$ is a small metric perturbation to the background curved spacetime metric $\bar{g}_{\mu\nu}$, which also can be defined as $h_{\mu\nu}\equiv \delta g_{\mu\nu}$.

A natural question in the decomposition of Eq.~\eqref{1} is how to distinguish the curved background $\bar{g}_{\mu\nu}$ from the perturbation $h_{\mu\nu}$ in a generic spacetime. Unlike the flat-spacetime case, this split is neither unique nor intrinsically meaningful; rather, it is defined operationally through a separation of scales \cite{Maggiore}. We identify $\bar{g}_{\mu\nu}$ with the slowly varying part of the metric on length scales much larger than the gravitational-wave wavelength, while $h_{\mu\nu}$ captures the rapidly varying, small-amplitude fluctuations. Consequently, the distinction is gauge-dependent and becomes well-defined only after an appropriate averaging over spacetime regions (or wavelengths).

This decomposition is essential for a consistent description of gravitational waves beyond flat spacetime. In particular, it permits the construction of an effective energy--momentum tensor from second-order perturbations, which acts as a source for the background geometry. Gravitational waves therefore propagate on $\bar{g}_{\mu\nu}$ while also inducing a backreaction on it, making this framework central to studies of GW propagation, energy transport, and their role in curved and cosmological spacetimes.\\

In the present work, we restrict attention to terms linear in $h_{\mu\nu}$. While the split $g_{\mu\nu}=\bar{g}_{\mu\nu}+h_{\mu\nu}$ is unambiguous on a Minkowski background, it is more delicate on a generic curved spacetime because both $\bar{g}_{\mu\nu}$ and $h_{\mu\nu}$ vary with position. We therefore work in the short-wavelength (high-frequency) regime, where the perturbation wavelength $\lambda$ is much smaller than the characteristic curvature scale $L$ of the background \cite{Maggiore, Mondal}. The combined weak-field and short-wavelength assumptions ($|h_{\mu\nu}|\ll 1$, $\lambda\ll L$) provide a consistent framework for gravitational-wave propagation in cosmological spacetimes: the perturbations behave as linear, rapidly varying modes on a slowly varying background geometry. In this limit, the leading contribution to the first-order correction of the Christoffel symbols, $\Gamma^{(1)}$, is dominated by the short-wavelength modes.

Accordingly, we neglect all higher-order terms $\mathcal{O}(h^2)$, thereby discarding the nonlinear contributions that would otherwise generate an effective energy--momentum tensor for gravitational waves and their backreaction on the background spacetime. We also set the first-order matter perturbations to zero, $T^{\rm m(1)}_{\mu\nu}=0$, so that the dynamics are governed solely by linearized geometric perturbations propagating on a fixed curved background. This \emph{weak-field, high-frequency approximation} is standard in analyses of gravitational-wave propagation in cosmological spacetimes.\\

The inverse metric $g^{\mu\nu}$ can be expressed as \cite{Maggiore, Mondal}
\ben
g^{\mu\nu}(x) = \bar{g}^{\mu\nu}(x) - h^{\mu\nu}(x) + \mathcal{O}(h^2).
\label{2}
\een
The Christoffel symbol ($\Gamma^{\mu}_{\nu\rho}$) with a linearized order perturbation is expressed as
\ben
\Gamma^{\mu}_{\nu\rho}
= \bar{\Gamma}^{\mu}_{\nu\rho}
+ \delta\Gamma^{\mu}_{\nu\rho},
\label{3}
\een
where $\bar{\Gamma}^{\mu}_{\nu\rho}$ is associated with the background metric $\bar{g}_{\mu\nu}$ and the linearly perturbed connection coefficient has the following form \cite{Maggiore},
\ben
\delta\Gamma^{\mu}_{\nu\rho}=\Gamma^{\mu(1)}_{\nu\rho}
= \frac{1}{2}\bar{g}^{\mu\sigma}
\left(
\bar{\nabla}_\nu h_{\rho\sigma}
+ \bar{\nabla}_\rho h_{\nu\sigma}
- \bar{\nabla}_\sigma h_{\nu\rho}
\right).
\label{4}
\een
It is to note that here, $\bar{\nabla}_\mu$ is the covariant derivative with respect to background metric $\bar{g}_{\mu\nu}$ and mention the condition of metric compatibility $\bar{\nabla}_\lambda \bar{g}_{\mu\nu}=0$. 

The Ricci tensor can be split into background, and linearly perturbed Ricci tensor as 
\ben
R_{\mu\nu} = \Big(\bar{R}_{\mu\nu} + R^{(1)}_{\mu\nu}\Big),
\label{5}
\een
where the linearized Ricci tensor is \cite{Maggiore}
\ben
R^{(1)}_{\mu\nu}
= \frac{1}{2}
\left(
\bar{\nabla}_\alpha \bar{\nabla}_\mu h^\alpha_{\nu}
+ \bar{\nabla}_\alpha \bar{\nabla}_\nu h^\alpha_{\ \mu}
- \bar{\Box} h_{\mu\nu}
- \bar{\nabla}_\mu \bar{\nabla}_\nu h
\right),
\label{6}
\een
where $h = h^\lambda_{\lambda}$ is the stress of the metric perturbation and $\bar{\Box} = \bar{g}^{\mu\nu}\bar{\nabla}_\mu\bar{\nabla}_\nu$. The notable thing is that $\bar{\nabla}_\nu \bar{\nabla}_\mu h= \bar{\nabla}_\nu \partial_\mu h=\Big(\partial_\nu\partial_\mu h-\bar{\Gamma}^\rho_{\nu\mu}\partial_\rho h \Big)$ is symmetric under the exchange $\mu\leftrightarrow\nu$ and therefore $R^{(1)}_{\mu\nu}$ is also symmetric \cite{Maggiore}.

For algebraic simplification, we have used the trace-reversed perturbation variable which can be defined as \cite{Maggiore, Mondal} 
\ben
\tilde{h}_{\mu\nu}
= h_{\mu\nu} - \frac{1}{2}\bar{g}_{\mu\nu} h ,
\label{7}
\een
with the trace of metric perturbation
\ben
\tilde{h} = -h .
\label{8}
\een
Therefore, we can write
\ben
h_{\mu\nu}=\tilde{h}_{\mu\nu}-\frac{1}{2}\bar{g}_{\mu\nu}\tilde{h} \nonumber \\
h^\alpha_\nu=\tilde{h}^\alpha_\nu-\frac{1}{2}\delta^\alpha_\nu \tilde{h} 
\label{9}
\een
In the present analysis, we work in the weak-field, high-frequency regime of linearized gravity. It is therefore convenient to introduce the trace-reversed perturbation, which removes redundant trace terms and simplifies the linearized Einstein equations. In this form the field equations become effectively diagonal, making the propagating degrees of freedom manifest and allowing the dynamics to be cast as a relativistic wave equation for metric perturbations (i.e., gravitational waves).

Since $R^{(1)}_{\mu\nu}$ is symmetric under $\mu\leftrightarrow\nu$, the corresponding second derivatives commute, so that
$\bar{\nabla}_\nu\bar{\nabla}_\mu\tilde{h}=\bar{\nabla}_\mu\bar{\nabla}_\nu\tilde{h}.$

For gravitational waves on a curved background, it is convenient to impose the Lorentz (Hilbert/harmonic/De Donder) gauge \cite{Maggiore, Maggiore1, Wald, Carroll, Schutz, Blau}, which partially fixes the coordinate freedom and removes unphysical components of the metric perturbation. With this choice, the linearized field equations take a simpler wave-like form. On a curved background, the gauge condition is \cite{Mondal}:
\ben
\bar{\nabla}_\mu \tilde{h}^{\mu\nu} = 0 
\label{10}
\een
Therefore, the linearly perturbed Ricci tensor becomes
\ben
R^{(1)}_{\mu\nu}
= \frac{1}{2}\Big[\bar{\nabla}_\alpha \bar{\nabla}_\mu \tilde{h}^\alpha_{\ \nu}+ \bar{\nabla}_\alpha \bar{\nabla}_\nu \tilde{h}^\alpha_{\ \mu}- \bar{\Box} \tilde{h}_{\mu\nu}\nonumber\\
+\frac{1}{2}\bar{g}_{\mu\nu} (\bar{\Box}\tilde{h})
\Big].
\label{11}
\een
and the linearized Ricci scalar can be written as
\ben
R^{(1)}=\frac{1}{2}\bar{\Box}\tilde{h}+\bar{R}^{\mu\nu}\tilde{h}_{\mu\nu}+\frac{1}{2}\bar{R}\tilde{h}
\label{12}
\een

According to the definition of the Einstein tensor $E_{\mu\nu}~\big(=R_{\mu\nu}-\frac{1}{2}g_{\mu\nu}R\big)$, where $E_{\mu\nu}$ denotes the Einstein tensor for notational convention. Now we can split the Einstein tensor into the two parts, as the background and the linearly perturbed parts, as follows
\ben
E_{\mu\nu}=\bar{E}_{\mu\nu}+E^{(1)}_{\mu\nu}. 
\label{13}
\een
The linearly perturbed form of the Einstein tensor can be written as
\ben
E^{(1)}_{\mu\nu}=R^{(1)}_{\mu\nu}-\frac{1}{2}\bar{g}_{\mu\nu}R^{(1)}-\frac{1}{2}h_{\mu\nu}\bar{R}.
\label{14}
\een
Thus, putting the values of $R^{(1)}_{\mu\nu}$ and $R^{(1)}$ from Eqs. (\ref{11}) and (\ref{12}) and trace-reversed term of the metric perturbation from Eqs. (\ref{8}) and (\ref{9}), we get
\ben
E^{(1)}_{\mu\nu}=\Big[&-&\frac{1}{2}\bar{\Box}\tilde{h}_{\mu\nu}+\bar{R}_{\alpha\mu\nu\beta}\tilde{h}^{\alpha\beta}+\frac{1}{2}\big(\bar{R}_{\lambda\mu}\tilde{h}^\lambda_\nu\nonumber\\&+&\bar{R}_{\lambda\nu}\tilde{h}^\lambda_\mu\big)-\frac{1}{2}\bar{g}_{\mu\nu}\big(\bar{R}^{\alpha\beta}\tilde{h}_{\alpha\beta}\big)-\frac{1}{2}\tilde{h}_{\mu\nu}\bar{R}\Big]\nonumber\\ 
\label{15}
\een
This is the final expression for the Einstein tensor to linear order in the perturbation, combining contributions from the linearized Einstein equations and the gravitational-wave equation.

\section{Scalar-Induced Gravitational Waves} \label{Sec.III}
In this section, we derive the linearized gravitational-wave equation in the presence of a k-essence scalar field, including both scalar and tensor perturbations.

\subsection{Linearized Perturbation in Scalar Field Model}

In the gravitational sector, we consider the k-essence is minimally coupled to GR \cite{Garriga, Picon, Picon1, Vikman, Malquarti, Chimento, Jorge, Chiba, gm1, gm2, gm3, Ganguly, Panda, Bamba, Ganguly1}:
\ben
S\equiv S_{\rm EH}+S_k=\int d^4x \,\sqrt{-g}\Big[\frac{R}{2\kappa^2}+\La(X,\phi)\Big],
\label{16}
\een
where $\kappa^2=8 \pi G$, $\La(X,\phi)$ is the non-canonical scalar Lagrangian, $g$ is the determinant of $g_{\mu\nu}$, and $X\equiv\frac{1}{2}g^{\mu\nu}\partial_\mu \phi \,\partial_\nu\phi$.

Although Eq.~\eqref{16} displays only the Einstein--Hilbert term plus the k-essence Lagrangian, our background cosmology also includes an additional matter/radiation sector (collectively denoted by $\psi_m$). Thus, the full setup assumed for the background is
\ben
S_{\rm total}=S_{\rm EH}+S_k+S_m[g_{\mu\nu},\psi_m],
\label{16b}
\een
so that the FLRW background metric is sourced by the \emph{total} stress-energy,
$\bar\rho=\bar\rho_\phi+\bar\rho_m$ and $\bar P=\bar P_\phi+\bar P_m$.
In the present work, we focus on the linear transverse-traceless tensor sector, and, consistently, we set first-order matter perturbations to zero, $T^{m(1)}_{\mu\nu}=0$.
For a minimally coupled scalar and a perfect-fluid matter sector, scalar/vector perturbations do not source the TT sector at linear order; therefore ordinary matter affects the GW propagation equation only through the background Friedmann evolution (Appendix~\ref{App:B}). At late times, we take matter to be pressureless ($\bar P_m\simeq 0$) and neglect radiation in the late-time analysis.

We begin with the scalar field $\phi$, which in general depends on both time and space, $\phi\equiv\phi(t,\mathbf{x})$. Introducing small perturbations about a homogeneous background, we decompose $\phi$ into a spatially independent background and a perturbation \cite{Maggiore1, Dodelson, Garriga}:
\ben
\phi(t,x)=\bar{\phi}(t)+\delta\phi(t,x) \label{17}
\een
where the background field $\bar{\phi}$ is only the function of time since our background universe is homogeneous in nature, and $|\delta\phi|<<\bar{\phi}$.

Here, the spacetime metric $g_{\mu\nu}$ and the k-essence scalar field $\phi$ are independent dynamical fields in the action. Accordingly, their perturbations $h_{\mu\nu}$ and $\delta\phi$ are defined independently and can be expanded about their respective background configurations, without assuming any direct algebraic relation between them.

Since we work at linear order, we retain only terms that are first order in $h_{\mu\nu}$ or $\delta\phi$ and neglect all quadratic and mixed contributions (e.g., $h\,\delta\phi$).

Therefore, the variation of the kinetic term can be derived as, 
\ben
&&\delta X= \Big[-\frac{1}{2}h^{\mu\nu}\partial_\mu\bar{\phi}.\partial_\nu\bar{\phi}\Big]+\bar{g}^{\mu\nu}\Big[\partial_\mu(\delta\phi).\partial_\nu\bar{\phi}\Big]\nonumber\\
\text{or,}&&\delta X= \delta X_h+\delta X_\phi \label {18}
\een
where, we have defined $\delta X_h=(-\frac{1}{2}h^{\mu\nu}\partial_\mu\bar{\phi}.\partial_\nu\bar{\phi})$ and $\delta X_\phi=\bar{g}^{\mu\nu}[\partial_\mu(\delta\phi).\partial_\nu\bar{\phi}]$.

The energy--momentum tensor of the k-essence scalar field (see Appendix~\ref{App:A}) is
\ben
T^{\phi}_{\mu\nu}=g_{\mu\nu}\,\La-\La_X\,\partial_\mu\phi\,\partial_\nu\phi,
\label{19}
\een
where $\La_X\equiv \partial\La/\partial X$, $\La_{X\phi}\equiv \partial^2\La/(\partial X\,\partial\phi)$, $\La_{XX}\equiv \partial^2\La/\partial X^2$, and $\La_{\phi}\equiv \partial\La/\partial\phi$.

Thus, the linearly perturbed energy-momentum tensor can be expressed as
\ben
\delta T^\phi_{\mu\nu}\equiv T^{\phi(1)}_{\mu\nu}= M_{\mu\nu}(h)+ S_{\mu\nu}(\delta \phi) \label{20}
\een
where, $M_{\mu\nu}(h)$ depends on metric perturbation term $h_{\mu\nu}$ in linear order and can be defined as
\ben
M_{\mu\nu}(h) \equiv [h_{\mu\nu}\La &+& \bar{g}_{\mu\nu}\La_X(\delta X_h)\nonumber\La\\ &-& \La_{XX}(\delta X_h)\partial_\mu\bar{\phi}.\partial_\nu\bar{\phi}] \label{21}
\een
and the second term in Eq.(\ref{20}), i.e., $S_{\mu\nu}(\delta\phi)$ depends on the scalar field perturbation term $\delta\phi$ and defined as 
\ben
S_{\mu\nu}(\delta\phi)&\equiv&[\bar{g}_{\mu\nu}\La_X(\delta X_\phi)+\bar{g}_{\mu\nu}\La_\phi(\delta \phi)\nonumber\\&-&\La_{XX}(\delta X_\phi)\partial_\mu\bar{\phi}.\partial_\nu\bar{\phi}- \La_{X\phi}(\delta \phi)\partial_\mu\bar{\phi}.\partial_\nu\bar{\phi}\nonumber\\&-& \La_X(\partial_\mu(\delta\phi).\partial_\nu\bar{\phi}+\partial_\nu(\delta\phi).\partial_\mu\bar{\phi})] \label{22}
\een
Note that Eqs.~(\ref{21}) and (\ref{22}) show that the two contributions to the inearized energy--momentum tensor in Eq.~(\ref{20}) depend separately on $h_{\mu\nu}$ and $\delta\phi$; there are no mixed terms proportional to $h_{\mu\nu}\,\delta\phi$ at this order.

Combining Eqs.~(\ref{14}) and (\ref{20}), \emph{the linearized Einstein equation} takes the form
\ben
\boxed{E^{(1)}_{\mu\nu}\equiv 8\pi G\,T_{\mu\nu}^{\phi(1)}\equiv 8\pi G\Big[M_{\mu\nu}(h)+S_{\mu\nu}(\delta\phi)\Big]},
\label{23}
\een
where we have set $T_{\mu\nu}^{m(1)}=0$.

Using Eqs.~(\ref{7})--(\ref{9}), $\delta X_h$ can be expressed in terms of the trace-reversed variable as
\ben
\delta X_h=-\delta X_{\tilde{h}}+\frac{1}{2}\bar{X}\,\tilde{h},
\label{24}
\een
where $\delta X_{\tilde{h}}\equiv \frac{1}{2}\tilde{h}^{\mu\nu}\partial_\mu\bar{\phi}\,\partial_\nu\bar{\phi}$ and $\bar{X}\equiv \frac{1}{2}\bar{g}^{\mu\nu}\partial_\mu\bar{\phi}\,\partial_\nu\bar{\phi}$. Accordingly, Eq.~(\ref{21}) can be rewritten in terms of the trace-reversed perturbation as
\ben
M_{\mu\nu}(\tilde{h})&=&\tilde{h}_{\mu\nu}\La-\bar{g}_{\mu\nu}\Big[\frac{1}{2}\tilde{h}\La+\La_X\Big(\frac{1}{2}\bar{X}\tilde{h}-\delta X_{\tilde{h}}\Big)\Big]\nonumber\\&+&\partial_\mu \bar{\phi} \partial_\nu \bar{\phi}\Big[\La_{XX}\Big(\delta X_{\tilde{h}}-\frac{1}{2}\bar{X}\tilde{h}\Big)\Big] \label{25}
\een

In what follows we treat the k-essence field as a purely homogeneous background, $\phi=\bar{\phi}(t)$, and set its intrinsic perturbation to zero, $\delta\phi=0$. At linear order (for a minimally coupled scalar), the transverse--traceless tensor sector evolves independently of scalar fluctuations \cite{Maggiore, Maggiore1, Wald, Carroll, Schutz, Blau}, so $\delta\phi$ does not enter the tensor evolution equation directly; the scalar field influences the dynamics only through background quantities in the stress--energy tensor. This choice is analogous to working in unitary gauge, where the scalar perturbation is absorbed into the metric variables \cite{Roy, Cheung, Gleyzes, Zumalacarregui}.

With $\delta\phi=0$ we have $\delta X_{\phi}=0$, and hence Eq.~(\ref{22}) implies $S_{\mu\nu}(\delta\phi)=0$. Therefore, Eqs.~(\ref{20}) and (\ref{25}) reduce to
\ben
T^{\phi(1)}_{\mu\nu}=M_{\mu\nu}(\tilde{h}) 
\label{26}
\een

Thus, at linear order, combining Eqs.~(\ref{15}), (\ref{23}), and (\ref{26}) yields \emph{the linearized Einstein equation}:
\ben
\boxed{-\frac{1}{2}\bar{\Box}\tilde{h}_{\mu\nu}+\tilde{W}_{\mu\nu}=8\pi G M_{\mu\nu}(\tilde{h})} \label{27}
\een
where
\ben
\tilde{W}_{\mu\nu}&=&\Big[\bar{R}_{\alpha\mu\nu\beta}\tilde{h}^{\alpha\beta}+\frac{1}{2}(\bar{R}_{\lambda\mu}\tilde{h}^\lambda_\nu+\bar{R}_{\lambda\nu}\tilde{h}^\lambda_\mu)\nonumber\\&-&\frac{1}{2}\bar{g}_{\mu\nu}(\bar{R}^{\alpha\beta}\tilde{h}_{\alpha\beta})-\frac{1}{2}\tilde{h}_{\mu\nu}\bar{R}\Big] \label{28}
\een

\subsection{Choice of Line Element with Scalar and Tensor Perturbation} \label{subsec:B}
In standard gravitational-wave theory on a flat or curved background, the metric perturbation $h_{\mu\nu}$ contains both gauge (coordinate) degrees of freedom and physical, radiative degrees of freedom. The transverse--traceless (TT) gauge \cite{Maggiore, Maggiore1, Wald, Carroll, Schutz, Blau} is particularly useful because it eliminates pure-gauge components and isolates the propagating tensor modes. Imposing appropriate gauge conditions and exploiting the residual coordinate freedom reduces the perturbation to its TT part, which is transverse to the direction of propagation and trace-free. Consequently, the gravitational wave is described by two independent polarization states, the plus $(+)$ and cross $(\times)$ modes, which encode the measurable tidal distortions. In the present theory, Eq.~(\ref{16}) shows that the inclusion of a k-essence scalar field $\phi$ in the Lagrangian $\La(X,\phi)$ introduces additional physical but non-propagating (non-radiative) degrees of freedom in $h_{\mu\nu}$. As a result, $h_{\mu\nu}$ cannot be made globally TT; instead, it can be decomposed into a radiative TT component that satisfies the TT conditions and complementary non-radiative pieces. The TT component obeys a wave equation in any frame, whereas the remaining components satisfy Poisson-type constraint equations, indicating that they are sourced and do not propagate as freely traveling radiation \cite{Dodelson, Eanna}.

We therefore consider scalar and tensor perturbations about an FLRW background ($\bar{g}_{\mu\nu}$) and work in the conformal Newtonian (longitudinal) gauge with the $(+,-,-,-)$ signature. In this gauge the scalar shift and scalar shear are set to zero ($B=E=0$), so that the scalar sector is fully characterized by two potentials. We then adopt the perturbed line element \cite{Das, Kohri, Li}
\ben
ds^2&&=a^2(\eta)\Big[(1+2\Phi)d\eta^2-\Big((1-2\Psi)\delta_{ij}\n &&+\frac{1}{2}h^{(TT)}_{ij}\Big)dx^i dx^j\Big] 
\label{29}
\een
where $\eta$ denotes conformal time and $a(\eta)$ is the scale factor. In the Newtonian gauge, $\Phi$ and $\Psi$ coincide with the gauge-invariant Bardeen potentials \cite{Maggiore1}. The tensor $h^{(TT)}_{ij}$ is the transverse--traceless perturbation describing gravitational waves, satisfying $\partial^i h^{(TT)}_{ij}=0$ and $h^{(TT)i}{}_i=0$. The Newtonian gauge is applicable to linear perturbations about an FLRW background for modes with nonzero wavenumber and is particularly convenient when vector modes are neglected so that the metric can be cleanly decomposed into scalar and TT tensor sectors. In cosmological (including k-essence) backgrounds, the Bardeen potentials are gauge-invariant combinations of scalar metric perturbations that encode the physical scalar degrees of freedom; the analogous potentials in our framework play the same role, removing gauge artifacts and capturing the scalar dynamics sourced by the k-essence field. A non-vanishing scalar anisotropic stress leads to a gravitational slip, $\Phi-\Psi\neq 0$, which has been studied in detail (see, e.g., \cite{Baumann}).

Several works adopt the same perturbed metric in Eq.~(\ref{29}) and include scalar and tensor perturbations up to second order \cite{Das, Kohri, Li}. In contrast, since we focus on linear perturbation theory, we retain only first-order scalar and tensor perturbations and neglect vector modes as well as anisotropic stress \cite{Li}.

Equation~(\ref{29}) implies the background conformal FLRW line element
\ben
d\bar{s}^2= \bar{g}_{\mu\nu} \, dx^\mu dx^\nu= a^2(\eta)\big[d\eta^2-\delta_{ij} \, dx^i dx^j\big] \label{30}
\een
and the corresponding nonvanishing perturbation components are obtained directly from Eq.~(\ref{29}):
\ben
&h_{00}=&2a^2(\eta)\Phi \nonumber\\
&h_{ij}=&-a^2(\eta)\Big[-2\Psi \delta_{ij}+\frac{1}{2}h^{(TT)}_{ij}\Big] \nonumber\\
&h_{0i}=&0 \label{31}
\een
and the trace of that metric perturbation: 
\ben
h=\bar{g}^{\mu\nu}h_{\mu\nu}=2(\Phi-3\Psi) 
\label{32}
\een

For algebraic convenience, we use Eqs.~(\ref{7}) and (\ref{8}) to express the perturbations in Eqs.~(\ref{31}) and (\ref{32}) in terms of the trace-reversed variable, yielding
\ben
&\tilde{h}_{00}&=\Big(h_{00}-\frac{1}{2}\bar{g}_{00}h\Big)=a^2(\Phi+3\Psi)\nonumber\\
&\tilde{h}_{ij}&=a^2(\Phi-\Psi)\delta_{ij}-\frac{a^2}{2}h^{(TT)}_{ij}\nonumber\\
&\tilde{h}&=-h=2(3\Psi-\Phi) \label{33}
\een

Since we work in conformal time $\eta$, we treat the homogeneous background scalar field as $\bar{\phi}=\bar{\phi}(\eta)$. Using Eqs.~(\ref{29}) and (\ref{33}), the background kinetic term and its perturbation can be written as
\ben
\bar{X}&=&\frac{(\bar{\phi}')^2}{2 a^2}\nonumber\\
\delta X_{\tilde{h}}&=&\bar{X}(\Phi+3\Psi) \label{34}
\een
where a prime denotes differentiation with respect to conformal time, $'\equiv \partial_\eta$.

Substituting Eqs.~(\ref{33}) and (\ref{34}) into Eq.~(\ref{25}), we obtain the nonvanishing components of $M_{\mu\nu}(\tilde{h})$ as
\ben
M_{00}(\tilde{h})&=&2a^2\Phi\Big[\La +\bar{X}\La _{X}+2\bar{X}^2\La _{XX}\Big]\nonumber\\
M_{ij}(\tilde{h})&=&2a^2\Big[\La\, \Psi-\bar{X}\La _X\,\Phi\Big]\delta_{ij}-\frac{a^2}{2}\La\, h^{(TT)}_{ij}\nonumber\\ \label{35}
\een

Similarly, inserting the nonzero background curvature tensors listed in Appendix~\ref{App:B} into Eq.~(\ref{28}), we obtain the nonvanishing components of the curvature contribution $\tilde{W}_{\mu\nu}$ appearing in Eq.~(\ref{27}) as
\ben
\tilde{W}_{00}&=&3\mathcal{H}'[3\Psi-\Phi]+12\mathcal{H}^2\Psi \nonumber\\
\tilde{W}_{ij}&=& \delta_{ij} \left[ \mathcal{H}'(7\Psi - 5\Phi) - 6\mathcal{H}^2(\Phi - \Psi) \right]\nonumber\\ &+& (2\mathcal{H}' + 3\mathcal{H}^2) h^{(TT)}_{ij}, \label{36}
\een
where $\mathcal{H}$ is the conformal Hubble parameter, which can be defined as $\mathcal{H}\equiv\big(\frac{a^\prime}{a}\big)$.

\section{Modified GWs Equation through a non-canonical Lagrangian} \label{Sec.IV}
In this section, we derive the modified gravitational-wave equation for a particular choice of non-canonical Lagrangian and then rewrite it in terms of the redshift $z$ for confronting the model with observations.

\subsection{Form of Modified GWs Equation}
Substituting Eqs.~(\ref{33}), (\ref{35}), and (\ref{34}) into Eq.~(\ref{27}), we obtain two independent components of the Einstein equations: the $00$ component and the $ij$ component. The $00$ component together with the scalar part of the $ij$ component describes the relation between the metric potentials $\Phi$ and $\Psi$ and governs their time evolution. By contrast, the tensor (transverse--traceless) part of the $ij$ component corresponds to tensor perturbations, i.e., gravitational waves.

As discussed earlier, in the presence of a k-essence background, the metric perturbations generically decompose into scalar, vector, and tensor sectors, with the scalar sector being sourced by the underlying scalar field and giving rise to additional, typically non-radiative modes. In contrast, the tensor sector, corresponding to the transverse, traceless (TT) part of the spatial metric perturbation, contains the true dynamical degrees of freedom associated with freely propagating gravitational waves \cite{Maggiore, Maggiore1, Wald, Carroll, Schutz, Blau}. This separation follows from the gauge-invariant decomposition of perturbations on a curved background, where only the TT tensor modes remain invariant under scalar and vector gauge transformations and satisfy a wave-like equation.

Physically, scalar and vector perturbations are either constrained by the field equations or decay with the cosmological expansion, and hence do not represent propagating radiative degrees of freedom. By contrast, the TT tensor modes propagate causally and carry energy and information across spacetime, making them the relevant observables for gravitational-wave physics. Therefore, to extract the physical propagation equation for gravitational waves from the perturbed Einstein equations, it is necessary and sufficient to project onto the TT sector of the spatial perturbations. This is achieved by applying the transverse--traceless projection operator, which systematically removes the non-dynamical (longitudinal and trace) components and isolates the tensor mode $h^{(TT)}_{ij}$ \cite{Maggiore, Maggiore1, Eanna}.

In the context of k-essence or more general modified-gravity scenarios, this procedure is particularly important because the scalar-field background can source additional metric perturbations that mix with the scalar sector but do not correspond to propagating gravitational radiation. The TT projection ensures that such non-radiative contributions are excluded, leaving a closed evolution equation for $h^{(TT)}_{ij}$ that captures the modified propagation of gravitational waves in the given background. Consequently, the dynamics of $h^{(TT)}_{ij}$ encodes the physically observable gravitational-wave signal, including possible deviations from general relativity induced by the k-essence field.

Accordingly, to obtain the gravitational-wave sector from the $ij$ component, we retain only the TT-surviving parts of  $\tilde{h}_{ij}$, $\tilde{W}_{ij}$, and $M_{ij}(\tilde{h})$ in Eqs.~(\ref{33}), (\ref{35}), and (\ref{36}), respectively. Acting with the TT projector \cite{Maggiore, Maggiore1, Eanna} then yields
\ben
\tilde{h}^{TT}_{ij}&=&-\frac{a^2}{2}h^{(TT)}_{ij}\nonumber\\
\tilde{W}^{TT}_{ij}&=&(2\mathcal{H}' + 3\mathcal{H}^2) h^{(TT)}_{ij}\nonumber\\
M^{TT}_{ij}(\tilde{h})&=&-\frac{a^2}{2}\La h^{(TT)}_{ij} \label{37}
\een
Under the TT projection, the $ij$ component of the Einstein equations reduces to
\ben
\boxed{-\frac{1}{2}\bar{\Box}\tilde{h}^{TT}_{ij}+\tilde{W}^{TT}_{ij}=8\pi G M^{TT}_{ij}(\tilde{h})} \label{38}
\een
Substituting Eq.~(\ref{37}) into Eq.~(\ref{38}), we obtain
\ben
\frac{1}{4}\bar{\Box}(a^2h^{(TT)}_{ij})&+&(2\mathcal{H}' + 3\mathcal{H}^2) h^{(TT)}_{ij}\nonumber\\ &+&4\pi G a^2  h^{(TT)}_{ij} \La(\bar{X},\bar{\phi})=0 \label{39}
\een
Applying the background d'Alembertian $\bar{\Box}=\bar{g}^{\mu\nu}\bar{\nabla}_\mu\bar{\nabla}_\nu$ in the TT ($ij$) sector, we obtain the simplified gravitational-wave equation as follows:
\ben
h^{(TT)''}_{ij} &+& 2\mathcal{H}h^{(TT)'}_{ij}-\nabla^2 h^{(TT)}_{ij}\nonumber\\ &+&\Big[8\mathcal{H}'+10\mathcal{H}^2+16\pi G a^2 \La(\bar{X},\bar{\phi})\Big]h^{(TT)}_{ij}=0 \nonumber\\ \label{40}
\een
Equation~(\ref{40}) is \emph{the modified gravitational-wave equation in the transverse-traceless (TT) sector} obtained in our setup, and it deviates from the standard GR result \cite{Maggiore}. The last bracketed term in Eq.~\eqref{40} is often referred to as \emph{mass-like} because it is proportional to $h^{(TT)}_{ij}$ itself (with no derivatives), i.e. it acts as a time-dependent effective potential in the TT wave equation. For clarity, in Sec.~\ref{Sub-Sec IV-B} we rewrite this contribution as a total background-induced potential $U^2(\eta)$ [Eq.~\eqref{47a}] and then define the \emph{k-essence-induced} effective mass $m_{\rm eff}^2$ as the scalar-sector part only [Eq.~\eqref{48}], such that $m_{\rm eff}^2\to 0$ when the k-essence Lagrangian is switched off. Using dimensional analysis in natural units ($c=\hbar=1$), we show that $U^2$ (and therefore $m_{\rm eff}^2$) has dimensions of mass squared (Appendix~\ref{App:C}).

\subsection{Specific Choice of Non-Canonical Lagrangian}
\label{Sub-Sec IV-B}
We now consider a specific choice of the non-canonical k-essence Lagrangian $\La(X,\phi)$ for our gravitational-wave analysis.

For the k-essence scalar field $\phi$, the pressure $P_{\phi}$, energy density $\rho_{\phi}$, equation-of-state (EoS) parameter $w_{\phi}$, and sound speed $c_s^2$ are given, respectively, by \cite{Picon, Picon1, Vikman, Malquarti, Chimento, Jorge, Ganguly, Ganguly1, Bamba, Padmanabhan}
\ben
&P_\phi=&\La(X,\phi)\nonumber\\
&\rho_\phi=&2X\La_X-\La \nonumber\\
&w_\phi=&\Big(\frac{P}{\rho}\Big)=\frac{\La(X,\phi)}{2X\La_X-\La}\nonumber\\
&c_s^2=&\frac{(\partial P/\partial X)}{(\partial \rho/\partial X)}=\frac{\La_X}{2X\La_{XX}+\La_X}. 
\label{41}
\een
From Eq.~\eqref{41}, we obtain a differential equation relating the sound speed $c_s^2$ to the Lagrangian $\La(X,\phi)$ (for $c_s^2\neq 0$):
\ben
2X\La_{XX}-\Big(\frac{1-c^2_s}{c^2_s}\Big)\La_X=0 
\label{42}
\een

Therefore, the general solution of Eq.~\eqref{42} is given by \cite{Sergijenko, Kunz, Kumar}:
\ben
\La(X,\phi)=U(\phi)X^n-V(\phi) \label{43}
\een
where $U(\phi)$ and $V(\phi)$ are arbitrary functions of $\phi$, and $n=(1+c_s^2)/(2c_s^2)$ is a constant.

For simplicity, we set $U(\phi)=1$, in which case the Lagrangian takes the form \cite{Dinda}
\ben
\La(X,\phi)=X^n-V(\phi). \label{44}
\een
This is the specific non-canonical k-essence Lagrangian adopted in our gravitational-wave analysis.

Here $V(\phi)$ plays the role of the potential for the scalar field $\phi$. Substituting Eq.~\eqref{44} into Eq.~\eqref{41}, we obtain
\ben
P_{\phi}=X^n-V(\phi),\nonumber\\
\rho_{\phi}=(2n-1)X^n+V(\phi),\nonumber\\
c_s^2=\frac{1}{2n-1}=\text{constant}.
\label{45}
\een
The sound speed must satisfy $0<c_s^2\leq 1$, which implies $n\geq 1$ \cite{Sergijenko, Kunz, Dinda}. From Eq.~\eqref{45}, one finds $c_s^2=1$ for $n=1$ (the quintessence limit), while for $n>1$ the sound speed decreases monotonically from unity as $n$ increases \cite{Dinda}. An important aspect of non-canonical scalar-field theories is the theoretical consistency of the perturbative sector: viable k-essence cosmologies must be free from both ghost and gradient instabilities. The no-ghost condition requires $\mathcal{L}_X>0$ \cite{Felice, Kobayashi, Chakraborty, Herrera},  which guarantees that the coefficient associated with the scalar perturbations proportional to $X\mathcal{L}_X/H^2$ do not carry negative kinetic energy \cite{Felice}. The absence of Laplacian instabilities further requires $c_s^2>0$ \cite{Felice}. For the power-law kinetic structure considered here, $c_s^2=1/(2n-1)$, so stability is automatically satisfied for $n>1/2$, while enforcing subluminal propagation, $c_s^2\le 1$, restricts the model to the physically relevant regime analyzed in this work.

Accordingly, the parameter region used in our observational analysis satisfies these stability conditions throughout the relevant stages of cosmic evolution. Moreover, since the tensor propagation speed remains exactly luminal, $c_{\rm gw}=1$, the model naturally evades the stringent multimessenger constraints from GW170817 \cite{Abbott}. In this framework, the non-canonical kinetic sector primarily affects the dispersive properties of gravitational-wave propagation through the effective tensor mass term, while preserving the causal structure and stability of the underlying theory.

For a homogeneous background in conformal time, the non-canonical k-essence Lagrangian becomes
$\La(\bar{X},\bar{\phi})=\Big(\frac{\bar{\phi}'^2}{2a^2}\Big)^n-V(\bar{\phi})$.
The corresponding background energy density $\bar{\rho}_{\phi}$, pressure $\bar{P}_{\phi}$, and equation-of-state parameter $\bar{w}_{\phi}$ then follow from Eq.~\eqref{45}:
\ben
&\bar{\rho}_\phi=&(2n-1)\bar{X}^n+V(\bar{\phi}) \nonumber\\
&\bar{P}_\phi=&\bar{X}^n-V(\bar{\phi})\nonumber\\
&\bar{w}_\phi=&\Big(\frac{\bar{P}_\phi}{\bar{\rho}_\phi}\Big)=\frac{\bar{X}^n-V(\bar{\phi})}{(2n-1)\bar{X}^n+V(\bar{\phi})}\label{46}
\een

Thus, at linear order, Eq.~\eqref{40} can be cast as a \emph{GR-like} tensor propagation equation plus an explicit k-essence correction, with the matter-sector perturbation set to zero ($T_{\mu\nu}^{m(1)}=0$):
\ben
\boxed{h^{(TT)''}_{ij} + 2\mathcal{H}h^{(TT)'}_{ij}-\nabla^2 h^{(TT)}_{ij}+m^2_{\rm eff} \, h^{(TT)}_{ij}=0} \nonumber\\ \label{47}
\een
To make the splitting transparent, we first introduce the total background-induced potential term (third bracket in Eq.~\eqref{40}),
\ben
U^2(\eta)\equiv 8\mathcal{H}'+10\mathcal{H}^2+16\pi G a^2\La(\bar{X},\bar{\phi})\,.
\label{47a}
\een
Using the conformal-time background Friedmann equations (Appendix~\ref{App:B}) and assuming a pressureless (dust) matter component, this can be written as
\ben
U^2(\eta)&&=16\pi G a^2(\eta)\Big[(\bar{\rho}-\bar{P})\Big]\n &&=16\pi G a^2(\eta)\Big[(\bar{\rho}_\phi-\bar{P}_\phi)+\bar{\rho}_m\Big].
\label{47b}
\een

In Eq.~\eqref{47}, we define $m_{\rm eff}^2$ as the k-essence-induced effective mass, chosen to denote \emph{only} the k-essence (scalar-sector) contribution, i.e. the deviation from the standard GR tensor equation on the same FLRW background,
\ben
m^2_{\rm eff}(\eta)&& \equiv U^2(\eta)-16\pi G a^2(\eta)\,\bar{\rho}_m(\eta)
\n &&=16\pi G a^2(\eta)\big(\bar{\rho}_\phi-\bar{P}_\phi\big)
\n &&=16\pi G a^2(\eta)\,(1-\bar{w}_\phi)\,\bar{\rho}_\phi\,.
\label{48}
\een
By construction, $m_{\rm eff}^2\to 0$ when the k-essence Lagrangian is switched off, $\La(\bar X,\bar\phi)\to 0$ (so that $\bar{\rho}_\phi=\bar P_\phi=0$), thereby recovering the standard GR tensor equation in the presence of matter \cite{Maggiore1}.

It is important to clarify the physical interpretation of the quantity $m_{\rm eff}^2$ appearing in Eq.~\eqref{48}. This term should not be interpreted as a fundamental graviton mass in the sense of Fierz--Pauli or dRGT massive-gravity theories~\cite{FierzPauli,dRhamGabadadze2010,dRhamGabadadzeTolley2011,HassanRosen2012}. The propagating tensor sector still contains only the standard massless spin-$2$ degrees of freedom of general relativity, i.e., the two transverse polarizations propagating at exactly luminal speed, $c_{\rm gw}=1$. Rather, the TT equation contains a background-induced, time-dependent potential term (summarized by $U^2$ in Eq.~\eqref{47a}); in Eq.~\eqref{48} we define $m_{\rm eff}^2$ to isolate the \emph{k-essence (scalar-sector)} part of this potential, such that $m_{\rm eff}^2\to 0$ when the k-essence Lagrangian is switched off. In this sense, $m_{\rm eff}^2$ provides a medium-induced dispersive correction to the tensor wave equation, modifying the dispersion relation without introducing additional tensor degrees of freedom or altering the causal structure of the theory. The observable consequence is therefore a frequency-dependent cumulative phase shift accumulated over cosmological distances, not a change in the fundamental GW propagation speed. This behavior is analogous to electromagnetic wave propagation in a plasma, where an effective mass-like contribution arises from interaction with the background medium without implying a genuine photon rest mass~\cite{Landau, Mendonca, Chen}. Finally, at linear order in cosmological perturbation theory the transverse--traceless tensor sector is gauge invariant under scalar and vector coordinate transformations; hence the TT-projected tensor evolution equation \eqref{37} is physically meaningful and free of residual scalar gauge ambiguities. The effective term $m_{\rm eff}^2$ defined here is therefore a genuine scalar-background-induced correction rather than a coordinate artifact.
\\

We now write the Fourier transform of the tensor perturbation $h^{(TT)}_{ij}$ as \cite{Maggiore, Maggiore1, Das, Assadullahi}:
\ben
h^{(TT)}_{ij}(\eta, \mathbf{x})=\sum_{A=+,\times}
\int \frac{d^3k}{(2\pi)^{\frac{3}{2}}} h_A(\eta,\mathbf{k}) e_{ij}^{(A)}(\mathbf{k}) e^{i\mathbf{k}\cdot\mathbf{x}} \nonumber\\ 
\label{49}
\een
where $h_A(\eta,\mathbf{k})$ denotes the Fourier mode of the gravitational-wave amplitude, $\mathbf{k}$ is the comoving wavevector, and $e_{ij}^{(A)}(\mathbf{k})$ is the (symmetric) transverse--traceless polarization tensor \cite{Das, Maggiore1}. The index $A$ labels the two polarization states, $A=+,\times$, and $e^{i\mathbf{k}\cdot\mathbf{x}}$ denotes the plane-wave basis. We emphasized that the normal-mode (Fourier) expansion in Eqs.~(\ref{49}) and (\ref{51}) is a purely classical plane-wave decomposition of the linearized tensor perturbation on an FLRW background. By spatial homogeneity and isotropy, plane waves labeled by the comoving wave vector $\mathbf{k}$ form a natural basis, and linearity implies that each transverse-traceless polarization evolves independently mode by mode \cite{Maggiore1, Weinberg, Mukhanov1, Mukhanov2, Dodelson, Eanna}. The coefficients $h_A(\eta,\mathbf{k})$ (or $\chi_A$) are therefore ordinary complex-valued classical amplitudes; they are not promoted to operators and no commutation relations are imposed (in contrast with the quantum-field-theoretic expansion with creation and annihilation operators) \cite{Birrell}.

Substituting Eq.~\eqref{49} into Eq.~\eqref{47}, we obtain the modified linearized gravitational-wave equation in Fourier space:
\ben
\boxed{h''_A + 2\mathcal{H}h'_A+(k^2+m^2_{\rm eff}) h_A=0} 
\label{50}
\een
Equation~(\ref{50}) gives the propagation of tensor perturbations on a k-essence--driven FLRW background. The modification should \emph{not} be viewed as coming from an explicit non-minimal tensor coupling; rather, it arises because tensor modes propagate on a time-dependent geometry sourced by a non-canonical scalar field. After transverse--traceless (TT) projection and an appropriate field redefinition, the background curvature and the k-essence stress--energy combine into an effective (generally dispersive) potential, in direct analogy with the usual cosmological term $a''/a$ \cite{Maggiore1}.

Throughout we take a homogeneous background field $\phi=\bar\phi(\eta)$ and work in unitary gauge, $\delta\phi=0$, so scalar fluctuations do not enter the tensor sector at linear order \cite{Roy, Cheung, Gleyzes, Zumalacarregui}. Gravitational waves therefore propagate purely as metric tensor perturbations \cite{Dodelson, Eanna, Maggiore, Maggiore1}, with the k-essence field influencing their evolution through the scalar-background contribution encoded in $m_{\rm eff}^2$ [Eq.~\eqref{48}]. No independent tensor source is generated at linear order.

Finally, Eq.~(\ref{50}) is obtained after fixing the gauge to isolate the physical tensor degrees of freedom: we adopt conformal Newtonian (longitudinal) gauge for the scalar sector and impose the TT conditions $\partial^i h_{ij}=0$ and $h^i{}_i=0$ on the spatial metric perturbation, leaving the two gravitational-wave polarizations $h^{(\rm TT)}_{ij}$.

We emphasize that this decoupling is not an artifact of unitary gauge: even if the scalar fluctuation $\delta\phi$ were kept explicit, it would not enter the tensor evolution equation at linear order. In first-order cosmological perturbation theory the scalar, vector, and tensor sectors evolve independently, and a minimally coupled scalar field carries no anisotropic stress capable of sourcing transverse--traceless metric perturbations. Consequently, the tensor mode $h^{(\rm TT)}_{ij}$ evolves independently of $\delta\phi$ in the linear regime. The k-essence field affects gravitational-wave propagation only indirectly through background quantities (e.g., $a(\eta)$ and $\bar\phi(\eta)$) that build the effective mass term $m_{\rm eff}^2$ in Eq.~(\ref{50}). A genuine coupling between $\delta\phi$ and tensor modes would arise only beyond linear order, or in theories where the scalar sector generates anisotropic stress.\\

We now discuss the propagation speed of gravitational waves. In effective-field-theory (EFT) approaches to late-time cosmology, in particular the EFT of dark energy/modified gravity, the quadratic action for tensor modes can be written schematically as $S_T\sim\int d\eta\,d^3x\,a^2\big[Q_T\,(h'_{ij})^2-c_{gw}^2\,(\partial_k h_{ij})^2\big]$, so that the coefficient of the $k^2$ term in the tensor equation of motion directly defines the GW propagation speed $c_{gw}$ \cite{Gubitosi, Bloomfield, BelliniSawicki, Gleyzes, Cheung, Maggiore1}. Therefore, any departure from unity in front of $k^2$ in Eq.~(\ref{50}) would correspond to $c_{gw}\neq c$. The joint observation of the binary neutron-star merger GW170817 \cite{Abbott} and the associated gamma-ray burst GRB~170817A \cite{Abbott1} imposes a remarkably tight bound on this deviation, $|c_{gw}-c|/c\lesssim 10^{-15}$, thereby excluding large classes of scalar--tensor and vector--tensor modifications of GR and strongly restricting EFT operators that predict $c_{gw}\neq 1$ \cite{Creminelli, Ezquiaga, Sakstein, Baker}.

However, our model automatically satisfies these bounds because the k-essence field is minimally coupled and the Lagrangian takes the form $\La(X,\phi)=X^n-V(\phi)$. In this case, the tensor sector retains the standard gradient term, so the coefficient of $k^2$ in Eq.~(\ref{50}) remains unity and the GW speed is luminal, $c_{gw}^2=1$. The only modification to GW propagation in our setup is therefore dispersive: it enters through the effective mass term $m_{\rm eff}^2$, rather than through a change in the fundamental propagation speed.\\

In Eq.~(\ref{50}), the term $2\mathcal{H}h'_A$ acts as a Hubble-friction term induced by the cosmic expansion, where the conformal Hubble parameter is $\mathcal{H}\equiv a'/a$. For GW propagation on an expanding FLRW background it is often convenient to absorb this damping by redefining the tensor amplitude, thereby eliminating the first-derivative term from the mode equation.

In this context, we introduce an auxiliary (rescaled) tensor mode $\chi_A(\eta,\mathbf{k})$ via \cite{Belgacem, Belgacem1, Maggiore1}
\ben
h_A(\eta,\mathbf{k})=\frac{1}{a(\eta)}\,\chi_A(\eta,\mathbf{k})\,, \label{51}
\een
where the factor $1/a(\eta)$ accounts for the cosmological dilution of the GW amplitude during propagation. With this field redefinition, the Hubble-friction term $2\mathcal{H}h'_A$ in Eq.~(\ref{50}) is absorbed into the evolution equation for $\chi_A(\eta,\mathbf{k})$, yielding a second-order mode equation without a first-derivative (damping) term \cite{Belgacem, Belgacem1, Maggiore1, Winitzki}.

By introducing this variable $\chi_A(\eta, \mathbf{k})$, Eq.(\ref{50}) becomes,
\ben
\chi''_A+\Big(k^2+m^2_{\rm eff}-\frac{a''}{a}\Big)\chi_A=0 \label{52}
\een
Thus, the rescaling \eqref{51} removes the first-derivative (Hubble-friction) term $2\mathcal{H}h'_A$ in Eq.~\eqref{50} and casts the tensor-mode dynamics into a simpler second-order equation \eqref{52} for $\chi_A$, with an effective time-dependent potential $a''/a$ and an additional dispersive contribution from the effective mass term $m_{\rm eff}^2$. This form is often more transparent for physical interpretation and analytic estimates of GW propagation on an expanding background. The term $m_{\rm eff}^2$ therefore encodes the deviation from the standard GR tensor-mode equation for cosmological perturbations \cite{Maggiore1}.\\

We now adopt the high-frequency (short-wavelength) approximation, usually called the {\it sub-horizon} (or {\it sub-Hubble}) limit \cite{Maggiore1, Belgacem, Belgacem1, Dodelson}. It corresponds to perturbations with physical wavelength much smaller than the horizon scale, $\lambda_{\rm phys}\ll R_H$, where $R_H\equiv a/\mathcal{H}$ is the (comoving) Hubble radius. For tensor modes on an FLRW background, this regime simplifies the evolution equation by separating the genuine wave propagation from the large-scale expansion effects that dominate on super-horizon scales (where tensor perturbations effectively freeze) \cite{Maggiore1}. 

For sub-horizon modes, we have $k\eta\gg 1$. Moreover, during both the matter-dominated era and the late-time dark-energy-dominated epoch, the quantity $a''/a$ is of order $\eta^{-2}$ (up to factors of order unity) \cite{Maggiore1, Belgacem, Belgacem1, Dodelson, Winitzki}. Hence, in the sub-horizon regime, one typically has $k^2\gg a''/a$, and the term $a''/a$ in Eq.~(\ref{52}) can be neglected compared to $k^2$.

Therefore, within this approximation, Eq.~\eqref{52} reduces to
\ben
\chi''_A+\Big(k^2+m^2_{\rm eff}\Big)\chi_A=0, \label{53}
\een
which is a modified propagation equation for the GW modes in a non-trivial cosmological background.

Equation~\eqref{53} is derived under this assumption: the expansion varies adiabatically over a GW period, so the modes behave as weakly perturbed plane waves. For the frequencies relevant to ground-based interferometers (LIGO/Virgo/KAGRA, $f\sim\mathcal{O}(10\text{--}10^3)\,\mathrm{Hz}$) and space-based missions (LISA, $f\sim\mathcal{O}(10^{-4}\text{--}10^{-1})\,\mathrm{Hz}$), one typically has $k^2\gg a''/a$ over most of the propagation history, which justifies reducing Eq.~\eqref{52} to Eq.~\eqref{53}. Accordingly, the results in this section apply to astrophysical (sub-horizon) signals, and not to primordial modes near horizon crossing or ultra-low-frequency waves probed by pulsar timing arrays, for which long-wavelength curvature corrections can be important.\\

In standard general relativity, the corresponding equation, $\chi''_A+k^2\chi_A=0$ \cite{Belgacem, Belgacem1, Maggiore1}, describes freely propagating massless tensor perturbations with a linear dispersion relation $\omega=k$ \cite{Belgacem}. The extra term $m^2_{\rm eff}$ in Eq.~\eqref{53} acts as an effective (time-dependent) contribution to the wave dynamics, originating from the coupling between tensor perturbations and the background k-essence scalar field $\bar\phi$, as well as from the background expansion, and therefore modifies the dispersion relation. As a consequence, different Fourier modes accumulate different phases during propagation, producing a frequency-dependent phase shift absent in standard GR. This effect can, in principle, be exploited observationally to probe the properties of dark energy through GW measurements.\\

Accordingly, Eq.~\eqref{53} implies the modified dispersion relation
$\omega_{\rm eff}^2=k^2+m^2_{\rm eff}$ \cite{Winitzki}. As discussed earlier, $m_{\rm eff}$ is not a fundamental graviton mass; rather, it acts as an effective (time-dependent) potential for the tensor modes, encoding background-induced corrections from the expansion history and the k-essence dynamics. Causality is preserved because the coefficient of the spatial-gradient term $k^2$ remains unchanged: the modification enters instead as a dispersive, frequency-dependent correction. As a result, the group velocity tends to the speed of light at high frequency, while small phase shifts can accumulate over cosmological propagation distances. In the high-frequency limit $k^2\gg |m^2_{\rm eff}|$ \cite{Winitzki} one may expand
$\omega_{\rm eff}\simeq k\big(1+\frac{m^2_{\rm eff}}{2k^2}\big)$. Thus, the cumulative phase shift $\Delta\Theta$ acquired during propagation can be written as
\ben
|\Delta\Theta|=\int \frac{m^2_{\rm eff}}{2k}\, d\eta.
\label{54}
\een
Although the GW propagation speed remains equal to the speed of light--as ensured by the unmodified coefficient of the $k^2$ term in Eq.~\eqref{50}, the presence of the $m^2_{\rm eff}$ term induces a non-trivial dispersion relation. This, in turn, leads to a frequency-dependent modification of GW propagation and hence to cumulative phase shifts over cosmological distances.

An important feature of the modified gravitational-wave propagation considered in this framework is the potential observational signature associated with the cumulative phase correction given by the effective tensor mass term. Although the quantity $m_{\rm eff}^{2}$ remains relatively small (an ultra-light physical scale) throughout the cosmological evolution, its contribution to the gravitational-wave phase may accumulate over very large propagation distances, potentially generating observable dispersive signatures in the detected waveform. In this form, the cumulative phase shift introduced in Eq. \eqref{54} depends both on the cosmological background evolution and on the wavelength of the propagating tensor mode, implying that long-wavelength gravitational waves traveling across cosmological distances may be especially sensitive to these corrections.

In this context, the dispersive contribution associated with $m_{\rm eff}^{2}$ may lead to small but non-negligible modifications in the phase evolution of gravitational-wave signals compared with the standard prediction of general relativity,  since the correction accumulates continuously during propagation. Moreover, the phase correction generated by the effective tensor mass differs qualitatively from a simple modification of the waveform amplitude, since it introduces a frequency-dependent propagation effect associated with the cosmological background itself. This feature could, in principle, allow future GW observations to distinguish the present scenario from standard general relativity through precision analyses of waveform propagation over cosmological distances.

\begin{figure}
    \centering
     % Subfigure (a)
    \begin{subfigure}[b]{0.48\textwidth}
        \centering
        \includegraphics[width=\textwidth]{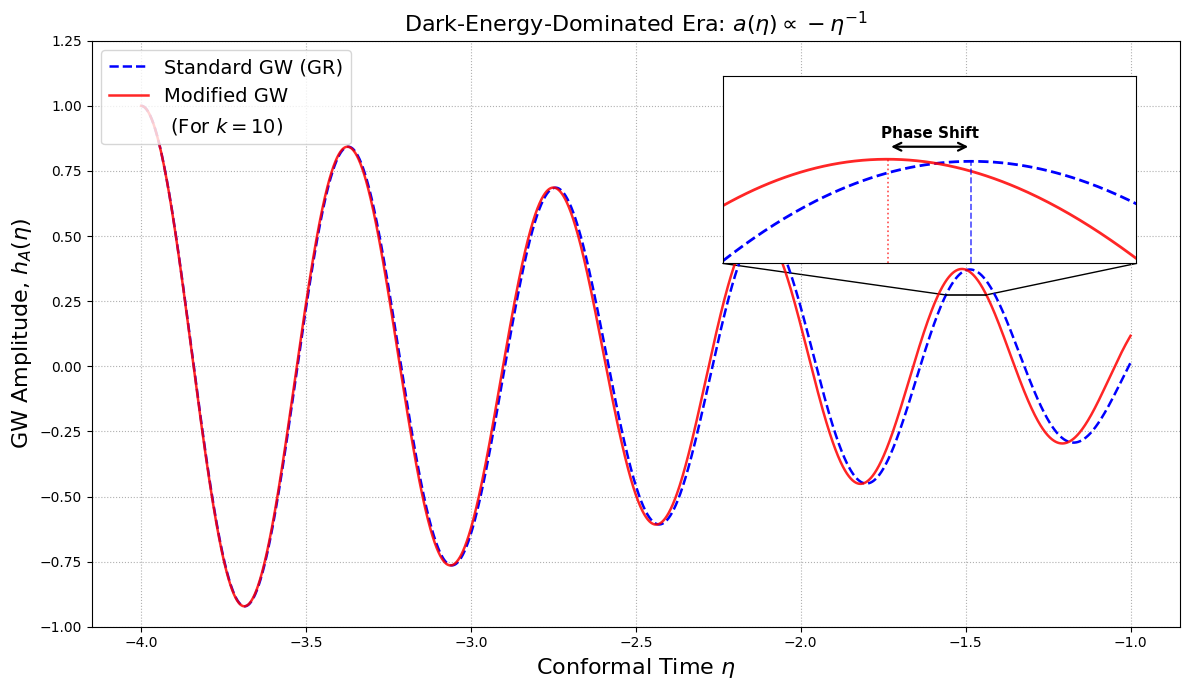}
        \caption{}
        \label{fig:1a}
    \end{subfigure}
    \hfill
    % Subfigure (b)
    \begin{subfigure}[b]{0.48\textwidth}
        \centering
        \includegraphics[width=\textwidth]{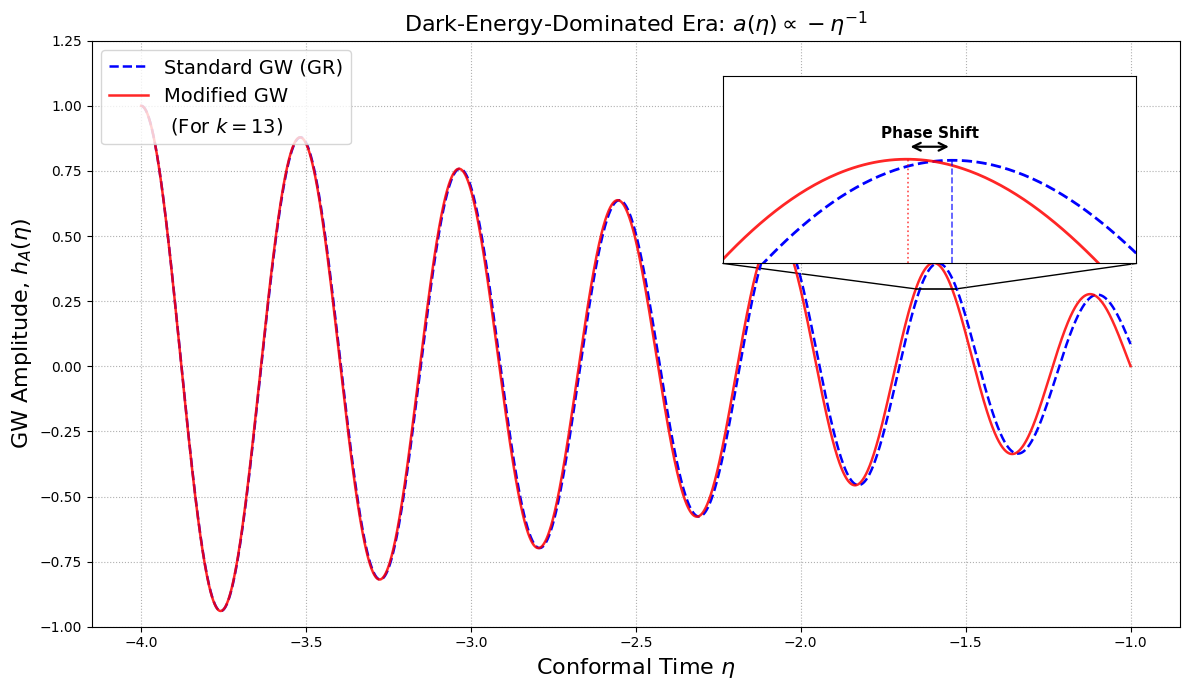}
        \caption{}
        \label{fig:1b}
    \end{subfigure}
    \hfill
    % Subfigure (c)
    \begin{subfigure}[b]{0.48\textwidth}
        \centering
        \includegraphics[width=\textwidth]{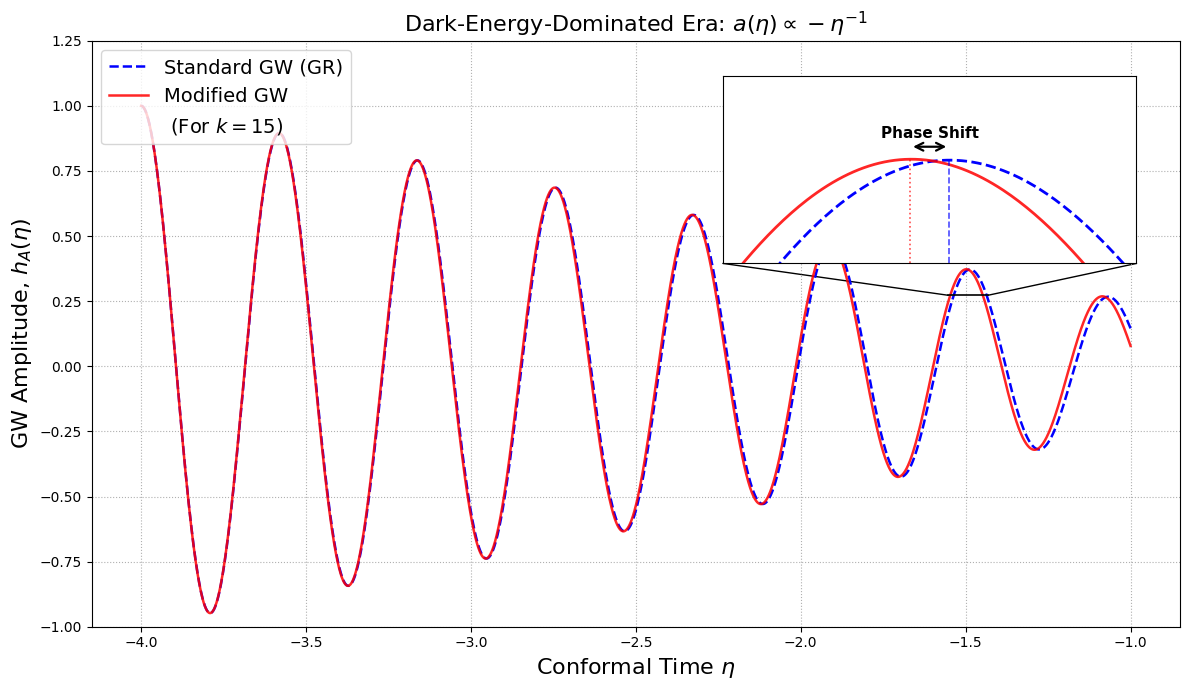}
        \caption{}
        \label{fig:1c}
    \end{subfigure} 
\caption{Comparison of the gravitational-wave amplitude $h_A(\eta)$ as a function of conformal time $\eta$ in a late-time, dark-energy-dominated background with $a(\eta)\propto -\eta^{-1}$ for $\eta<0$. The solid red curve shows the modified GW propagation governed by Eq.~(\ref{50}), with the effective mass term $m_{\rm eff}^2$ defined in Eq.~(\ref{48}); the dashed blue curve shows the standard GR propagation obtained by setting $m_{\rm eff}^2=0$. Panels (\ref{fig:1a})--(\ref{fig:1c}) correspond to representative Fourier wavenumbers $k=10$, $k=13$, and $k=15$, respectively. In each case, the modified waveform develops a cumulative phase shift relative to the GR prediction. The chosen values of $k$ are purely illustrative and are used as toy-model examples to show the phase-shift difference between the standard GR case and the modified propagation when the $m_{\rm eff}^{2}$ term is included.}
    \label{Fig1}
\end{figure}

To make the phase-shift induced by the effective mass term $m_{\rm eff}^{2}$ in Eq.~(\ref{50}) transparent, we work with an idealized late-time dark-energy-dominated background, rather than the best-fit cosmological parameters obtained from the data analysis in Sec.~\ref{sec.VI}. In the illustrative setup of Fig.~\ref{Fig1}, we identify the background scalar-field equation-of-state parameter with the dark-energy equation-of-state parameter, $\bar{w}_\phi=w_{\rm de}$, consistent with the interpretation adopted in Sec.~\ref{sec.VI}, and we set $\bar{w}_\phi=-1$. We further take $a(\eta)\propto -\eta^{-1}$, for which $\mathcal{H}=-1/\eta$ \cite{Dodelson, Mukhanov2}, and for convenience we set $8\pi G=1$ in Eq.~(\ref{48}). To illustrate the phase shift at the level of individual Fourier modes, we show three representative wavenumbers, $k=10$, $13$, and $15$, in panels (\ref{fig:1a}), (\ref{fig:1b}), and (\ref{fig:1c}) of Fig.~\ref{Fig1}, respectively. These values are not inferred from data or theoretically preferred; they are chosen solely to make the cumulative phase shift clearly visible.

Figure~\ref{Fig1} compares the gravitational-wave evolution in standard GR with that in the k-essence-modified case. The modified signal gradually accumulates a phase offset (dephases) relative to the GR waveform. This behavior is driven by the additional $m_{\rm eff}^2$ contribution in Eq.~(\ref{50}), which alters the tensor-mode dispersion relation. For the three representative choices of $k$, the phase difference is largest at smaller $k$ and becomes weaker as $k$ increases. This trend is consistent with the approximate expression $\omega_{\rm eff}\simeq k\!\left(1+\frac{m_{\rm eff}^2}{2k^2}\right)$ in the regime $|m_{\rm eff}^2|\ll k^2$ \cite{Winitzki} discussed above. Accordingly, Fig.~\ref{Fig1} should be viewed as a toy illustration of how the background-dependent term $m_{\rm eff}^2$ can generate a cumulative phase shift during GW propagation, rather than as an observational prediction or a parameter reconstruction for the k-essence model.\\

A more detailed numerical study, including realistic waveform templates and detector sensitivities, would be necessary in order to quantify the precise observational detectability of the predicted dispersive effects. Such an analysis lies beyond the scope of the present work and will be explored in future investigations. In addition, future high-precision gravitational-wave observations could provide an important observational window into the cosmological dynamics of the underlying k-essence sector. In particular, observations involving sources located at high redshifts may enhance the cumulative effect and therefore improve the possibility of constraining the effective tensor mass evolution.\\

Before proceeding further, we clarify the scope of field interactions considered in this work. In astrophysical environments, such as core-collapse supernovae \cite{Jones1} and compact binary systems \cite{Jones2}, gravitational waves can couple to electromagnetic and acoustic fields and may generate secondary radiation \cite{Barrett, Jones3, Jones4, Jones5, Jones6}. Such effects are typically associated with media that support anisotropic stresses and/or nonlinear couplings \cite{Jones7}. Here, by contrast, we study the linear propagation of transverse--traceless tensor perturbations on a homogeneous and isotropic FLRW background sourced by a minimally coupled k-essence scalar field. In linear cosmological perturbation theory, the scalar, vector, and tensor sectors evolve independently \cite{Mukhanov2, Weinberg, Maggiore1}, and at first order the tensor modes are not sourced by electromagnetic or scalar perturbations. In particular, for a homogeneous cosmological background one has $\langle F_{\mu\nu} \rangle = 0$, so no electromagnetic anisotropic stress is generated to couple to the tensor sector. Gravitational waves therefore propagate as free tensor modes, with their dynamics modified only through background quantities entering the effective field equations.

On cosmological scales, and within the high-frequency (sub-horizon) approximation adopted here, the decoupling of the radiative tensor sector from scalar and vector perturbations is well justified \cite{Mukhanov2, Weinberg, Maggiore1, Dodelson}. In this regime, the leading modification to gravitational-wave propagation is controlled by the time-dependent k-essence background and is conveniently encoded in a redshift-dependent effective mass term \cite{Nishizawa, Belgacem, Belgacem1}. While electromagnetic mixing may be relevant in localized astrophysical events \cite{Jones1, Jones2, Barrett, Jones3, Jones4, Jones5, Jones6, Jones7}, its contribution to the large-scale, linear propagation of cosmological GWs is expected to be negligible compared to the geometric effects induced by the dark-sector background \cite{Mukhanov2, Weinberg, Maggiore1}.

Accordingly, our goal is to isolate the impact of the k-essence dark sector on the gravitational-wave dispersion relation within a minimal and controlled setup. Extensions that include electromagnetic fields, non-minimal couplings, or higher-order perturbative effects could introduce additional mixing phenomena and are an interesting direction for future work, but are beyond the scope of the present analysis.

\subsection{Formulations in Redshift $(z)$ Space}
In this subsection, we transform the dependence of the effective mass term $m_{\rm eff}^2$ in Eq.~(\ref{48}) from conformal time $\eta$ to redshift $z$.\\

We use the following relations \cite{Maggiore1, Belgacem, Belgacem1, Dodelson}:
\ben
\frac{a(\eta)}{a_0}&=&\frac{1}{1+z}\nonumber\\
\mathcal{H}&=&-\frac{1}{1+z}\frac{dz}{d\eta}\label{55}
\een
where $a_0$ is the present value of the scale factor.

At late times, the contribution of radiation is negligible compared to that of dark matter and dark energy \cite{Aghanim}. Therefore, imposing covariant conservation of the total background energy--momentum tensor (see Appendix~\ref{App:B}, Eq.~(\ref{B10})), $\bar{\nabla}_\mu\bar{T}^{\mu\nu}=0$, we obtain the following continuity equations:
\ben
\bar{\rho}'_\phi +3\mathcal{H}(\bar{\rho}_\phi +\bar{P}_\phi)=0\nonumber\\
\bar{\rho}'_m +3\mathcal{H}(\bar{\rho}_m +\bar{P}_m)=0 \label{56}
\een

If the background scalar field $\bar{\phi}$ accounts for the dark-energy sector, we may identify
$\Omega_\phi\equiv\Omega_{\rm de}$ and $w_\phi\equiv w_{\rm de}$, where $\Omega_{\rm de}$ and $w_{\rm de}$ denote the dark-energy density parameter and equation-of-state (EoS) parameter, respectively \cite{Dodelson,Aghanim}. In the standard $\Lambda$CDM model, dark energy is a cosmological constant $\Lambda$ with fixed EoS $w_{\rm de}=-1$. A widely used minimal extension is the $w$CDM model, in which the dark-energy EoS is promoted to a constant free parameter $w_{\rm de}\neq -1$. More general phenomenological extensions allow a redshift-dependent EoS, e.g. the CPL form $w(z)=w_0+w_a\,z/(1+z)$ \cite{Chevallier,Linder}. In our notation, this corresponds to a $w_\phi$CDM description.

Cold dark matter is non-relativistic at late times, with a tiny velocity dispersion, so its background pressure is negligible compared to its energy density, $\bar{P}_m\simeq 0$ (i.e. $w_m\equiv \bar{P}_m/\bar{\rho}_m\approx 0$), and the matter sector can be treated as dust \cite{Dodelson} (considering for our case). With $\bar{P}_m=0$, Eq.~(\ref{56}) yields
\ben
\bar{\rho}_m(z)&=&\bar{\rho}_{m,0}(1+z)^3,\nonumber\\
\bar{\rho}_\phi(z)&=&\bar{\rho}_{\phi,0}(1+z)^{3(1+\bar{w}_\phi)}
\label{57}
\een
where $\bar{\rho}_{\phi,0}$ and $\bar{\rho}_{m,0}$ are the present-day energy densities of the scalar field and matter, respectively. 

Using Eqs.~(\ref{48}) and (\ref{57}), we can now express $m_{\rm eff}^2$ as a function of redshift $z$ as
\ben
m^2_{\rm eff}(z)&=&\frac{6 H^2(z)}{(1+z)^2}\,(1-\bar{w}_\phi)\,\Omega_\phi(z)\nonumber\\
\text{or,}\nonumber\\ m^2_{\rm eff}(z)&=&\frac{6 H_0^2}{(1+z)^2}\,(1-\bar{w}_\phi)\,\Omega_{\phi,0}(1+z)^{3(1+\bar{w}_\phi)}\,\nonumber\\
\label{58}
\een
where $H_0$ is the present day Hubble constant and $\Omega_{\phi}\equiv\frac{\kappa^2 \bar{\rho}_{\phi}}{3H^2}$ (with $\Omega_{\phi,0}\equiv\frac{\kappa^2 \bar{\rho}_{\phi,0}}{3H^2_0}$) is the scalar-field density parameter. We have also used the relation between the conformal Hubble parameter $\mathcal{H}$ and the standard (physical) Hubble parameter $H$, i.e. $\mathcal{H}=aH$, with the scale factor $a=\frac{1}{(1+z)}$.

Equation~(\ref{58}) shows that the effective mass term $m_{\rm eff}^2$ controlling tensor-mode propagation is sourced only by the \emph{scalar} background. Using Eq.~(\ref{48}), one finds that $m_{\rm eff}^2$ is set by the scalar-field energy density $\bar{\rho}_\phi$ and pressure $\bar{P}_\phi$. Moreover, since $m_{\rm eff}^2$ depends on the scalar-field equation-of-state parameter $\bar{w}_\phi$ [see Eqs.~(\ref{48}) and (\ref{58})], different realizations of dark energy, and, in our setup, different choices of non-canonical Lagrangians $\La(X,\phi)$, lead to different redshift evolutions of $m_{\rm eff}^2(z)$ and, consequently, to distinct dispersive phase corrections for gravitational waves. By construction, $m_{\rm eff}^2(z)\to 0$ when the k-essence sector is switched off.\\

We now discuss the gravitational-wave luminosity distance $D_L^{\rm GW}(z)$. In an FLRW spacetime, photons and gravitons propagate on the same background geometry, and in GR both follow null geodesics; consequently, the luminosity distance $D_L(z)$ is determined solely by the expansion history. In many modified-gravity scenarios, a mismatch between $D_L^{\rm GW}$ and the electromagnetic luminosity distance $D_L^{\rm EM}$ can arise if the GW propagation speed differs from unity, $c_{\rm gw}\neq 1$, and/or if an anomalous friction term changes the amplitude damping during propagation \cite{Belgacem, Belgacem1, Garoffolo, Garoffolo1}. In our case, Eqs.~(\ref{47}) and (\ref{50}) show that the coefficient of friction term $h_A'(\eta, \mathbf{k})$ remains exactly $2\mathcal{H}$ and the propagation speed is luminal, $c_{\rm gw}=1$, so the GW amplitude decays as $h_A\propto a^{-1}$, exactly as in GR. Although the k-essence background induces the effective mass term $m_{\rm eff}^2$ in Eqs.~(\ref{47}) and (\ref{50}), it modifies only the dispersion relation (phase/frequency evolution) and does not affect the amplitude damping. Therefore, in this framework, we have $D_L^{\rm GW}(z)=D_L^{\rm EM}(z)$. This does \emph{not} render the GW data redundant. Standard-siren measurements provide \emph{direct} luminosity distances that are independent of the cosmic distance ladder and are affected by systematics that differ from those of supernovae and BAO. Moreover, whereas CC data constrain the local expansion rate $H(z)$, GW (and SN) distances probe the \emph{integrated} expansion history through $\int dz/H(z)$; combining differential and integral observables helps to break parameter degeneracies in the k-essence sector (e.g., among $H_0$, $\Omega_{m,0}$, and the scalar-field parameters governing $w_\phi(z)$). Finally, although our current likelihood uses only the distance sector (consistent with luminal propagation), the same k-essence background can generically induce dispersive phase corrections via $m_{\rm eff}^2(z)$; incorporating waveform-level propagation tests; therefore, we defer to future work. With these clarifications, we may define (see, e.g., Ref.~\cite{Weinberg})
\ben
D_L^{GW}(z)=c(1+z)\int_0^z \frac{dz_1}{H(z_1)} \label{59}
\een
where $c$ is the speed of light in unit of $km/s$. Now taking the derivative concerning $z$ in Eq. (\ref{59}), we can get the following differential equation:
\ben
\frac{dD_L^{GW}(z)}{dz}=\frac{D_L^{GW}(z)}{(1+z)}+\frac{c(1+z)}{H(z)}. \label{60}
\een

Neglecting the radiation component at late times, and assuming a spatially flat universe, we have
$\Omega_\phi(z)+\Omega_m(z)=1$. Using this condition in the Friedmann equation (Appendix~\ref{App:B}, Eq.~(\ref{B11})), the evolution of the Hubble parameter can be written as
\ben
\frac{dH}{dz}&=&\frac{3H(z)}{2(1+z)}\Big[1+w_\phi(z)\,\Omega_\phi(z)\Big].
\label{61}
\een
Here, we use the redshift conformal-time relation $\frac{dz}{d\eta}=-H(z),$ which follows from $z=a^{-1}-1$ together with $a'=a^2H$ (and the convention $a_0=1$, so $a=1/(1+z)$).

The evolution of the scalar-field density parameter $\Omega_\phi$ follows from its definition together with the continuity equations, and can be written as
\ben
\frac{d\Omega_\phi}{dz}&=&\frac{3\Omega_\phi}{1+z}\left(1+w_\phi\right)-\frac{2\Omega_\phi}{H}\frac{dH}{dz}.\label{62}
\een
Combining Eqs.~(\ref{61}) and (\ref{62}) gives the equivalent compact form
\ben
\frac{d\Omega_\phi}{dz}=\frac{3\,w_\phi(z)}{1+z}\,\Omega_\phi\left(1-\Omega_\phi\right),
\label{63}
\een
which is convenient for numerical integration.\\

\subsection{Evolution Equation for the Scalar Field Model}
Several cosmological studies have explored non-canonical scalar-field Lagrangians with a non-vanishing potential. In particular, Refs.~\cite{Bose, Santiago, Fang, Mukhanov} considered quadratic potentials in non-canonical frameworks, in which the potential energy can play a central role in driving inflation in the early universe. For canonical scalar fields, a quadratic potential arises naturally from effective field theory considerations and has been widely employed in modeling inflation, dark energy, and dark matter.

In this work, we adopt the quadratic potential
$V(\bar{\phi})=\frac{1}{2}m^2\bar{\phi}^2$ \cite{Bose, Santiago, Fang, Mukhanov}, where $m$ is a free, positive, real parameter. This choice introduces a characteristic mass scale and leads to coherent oscillations of the field; depending on the kinetic structure of the model, the averaged dynamics may effectively behave as pressureless matter or yield accelerated expansion \cite{Unnikrishnan}. When combined with a non-trivial kinetic term, the setup remains analytically tractable while exhibiting rich dynamics, making it well-suited to explore the interplay between kinetic and potential contributions to cosmic evolution.

This quadratic potential ($V(\phi)$) provides a simple and convenient framework for studying the evolution of the k-essence background together with the associated effects on gravitational-wave propagation. Nevertheless, the appearance of the effective tensor mass term is not exclusively tied to the quadratic form of $V(\phi)$. Rather, the modified dispersive behavior of gravitational waves originates primarily from the non-canonical kinetic structure of the k-essence sector and its coupling to the evolving cosmological background.

We therefore expect qualitatively similar gravitational-wave propagation effects to arise for other classes of scalar-field potentials, including exponential, inverse power-law, and plateau-type potentials commonly studied in dark-energy and inflationary scenarios. While the detailed redshift evolution of the effective tensor mass $m_{\rm eff}(z)$ will depend on the background dynamics associated with a given potential, the mechanism responsible for the frequency-dependent phase corrections remains the same, provided the non-canonical kinetic sector plays a non-negligible role in the cosmic evolution.

Different choices of $V(\phi)$ can modify the late-time evolution of the Hubble parameter and the scalar-field energy density, thereby changing both the amplitude and the redshift dependence of the dispersive corrections in the gravitational-wave sector. Consequently, future analyses with alternative potentials may reveal additional phenomenological signatures and could improve the model's compatibility with cosmological observations. A detailed comparative study of different scalar-field potentials is left for future work.\\

Using the non-canonical Lagrangian $\La(X,\phi)$ introduced in Eq.~(\ref{44}), the equation of motion (EoM) for the homogeneous background scalar field $\bar{\phi}$ reads
\ben
\bar{\phi}''+\frac{2(2-n)}{2n-1}\mathcal{H}\bar{\phi}'+\frac{2^{n-1}}{n(2n-1)}a^{2n}\bar{\phi}'^{(2-2n)}\Big(\frac{dV}{d\bar{\phi}}\Big)=0 \n
\label{64}
\een
where primes denote derivatives with respect to conformal time $\eta$. We introduce the field ``velocity''
\ben
\bar{\phi}' \equiv \frac{d\bar{\phi}}{d\eta}=v_\phi,\nonumber\\
\bar{\phi}'' \equiv \frac{dv_\phi}{d\eta},
\label{65}
\een
so that $v_\phi$ is the conformal-time derivative of $\bar{\phi}$. Moreover, from Eq.~(\ref{55}) we have $d\eta/dz=-1/H(z)$, and hence
$\frac{dz}{d\eta}=-H(z)$.

Since the redshift $z$ is the primary observational variable (rather than the conformal time $\eta$), it is useful to rewrite the background evolution equations in terms of $z$. We have already expressed the evolution equations for the Hubble parameter $H$ and the scalar-field density parameter $\Omega_\phi$ as functions of $z$ in Eqs.~(\ref{61}) and (\ref{62}). Using $d\eta/dz=-1/H(z)$, Eq.~(\ref{64}) can be recast as
\ben
\frac{dv_\phi}{dz}&=&\frac{2(2-n)}{2n-1}\frac{v_\phi}{(1+z)}\nonumber\\
&+&\frac{2^{n-1}}{n(2n-1)}\frac{v_\phi^{2-2n}}{H(z)\,(1+z)^{2n}}\Big(\frac{dV}{d\bar{\phi}}\Big),\label{66}
\een
and the field itself obeys
$\frac{d\bar{\phi}}{dz}=\frac{d\bar{\phi}}{d\eta}\frac{d\eta}{dz}=-\frac{v_\phi}{H(z)}$,
where we used the definition in Eq.~(\ref{65}).\\

Table~\ref{tab1} summarizes, for comparison, the leading tensor-propagation properties of several representative frameworks. In Table~\ref{tab1}, GR predicts two massless transverse-traceless tensor modes propagating at exactly luminal speed. Massive-gravity models can introduce a fundamental graviton mass and typically additional propagating degrees of freedom, leading to modified dispersion and, in general, extra polarizations. In scalar–tensor theories such as Horndeski gravity, non-minimal interactions may modify the tensor kinetic sector, allowing for model-dependent changes in $c_{\rm gw}$ and/or additional effective tensor interactions. In contrast, the minimally coupled k-essence scenario studied here preserves $c_{\rm gw}=1$ and introduces no extra polarizations; the only modification is a background-induced effective tensor potential that yields dispersive (phase) corrections without altering the underlying tensor degrees of freedom.

\begin{table*}[t]
    \centering
    \renewcommand{\arraystretch}{1.3}
    \caption{Comparison of tensor-propagation properties in different gravitational frameworks.}
    \label{tab1}
\begin{tabular}{c|c|c|c}
\hline\hline
{\bf Theory} & {\bf $c_{\rm gw}$} & {\bf Tensor mass} & {\bf Extra polarizations} \\
\hline
GR & $1$ & No & No \\
Massive gravity & Model-dependent & Fundamental & Yes \\
Horndeski & Model-dependent & Possible & Possible \\
Present work (minimally coupled k-essence) & $1$ & Effective/background-induced & No\\
\hline\hline
\end{tabular}
\end{table*}

\section{Methodology and data analysis with model fitting} \label{Sec.V}
In this section, we determine a set of present-day cosmological and field-theoretic quantities that characterize the background evolution: the Hubble constant $H_0$, the present matter density parameter $\Omega_{m,0}$, the scalar-field velocity $v_{\phi,0}$, the background field amplitude $\bar{\phi}_0$, and the kinetic (power-law) index $n$. These quantities are obtained by numerically integrating Eqs.~(\ref{61})--(\ref{63}) together with Eq.~(\ref{66}), subject to appropriate initial conditions. Among them, $\bar{\phi}_0$, $v_{\phi,0}$, and $n$ are model-specific parameters. The field amplitude $\bar{\phi}_0$ sets the overall scale of the potential and therefore fixes the normalization of the scalar-field mass via $m\propto \sqrt{V_0}/\bar{\phi}_0$, where $V_0\equiv V(\bar{\phi}_0)$ denotes the present-day ($z=0$) value of the potential. For sufficiently small $v_{\phi,0}$, the kinetic contribution is subdominant to $V(\bar{\phi}_0)$, so the background scalar-field equation-of-state approaches $\bar{w}_{\phi,0}\to -1$, consistent with the $\Lambda$CDM limit. We work in natural units ($c=\hbar=1$), so all quantities are expressed in powers of energy. In this convention, a scalar field has mass (energy) dimension, hence $\bar{\phi}_0$ is measured in energy units (typically eV) \cite{Ganguly1}. The velocity $v_{\phi,0}$, defined in Eq.~(\ref{65}) as a derivative with respect to conformal time $\eta$ (which has dimension of inverse energy), therefore has units of energy squared, i.e. $(\mathrm{eV})^2$. Finally, $n$ is dimensionless.\\

We then confront the model predictions with a compilation of observational data, including Cosmic Chronometer (CC) measurements \cite{Jimenez, zhang, Jimenez1, Simon, Stern, Moresco, Abbott2, Metin, Eric, Julian, Betoule}, Type Ia supernovae from the Pantheon+SHOES sample \cite{Brout, Scolnic}, Baryon Acoustic Oscillation (BAO) data \cite{Alam, Eisenstein1, Percival, Glazebrook, Hu, Eisenstein2, Dawson, DESI}, and the Gravitational-Wave Transient Catalogs (GWTC), namely GWTC-2.1 \cite{GWTC2.1}, GWTC-3 \cite{GWTC3}, and GWTC-4 \cite{GWTC4, Abac, Abac2}. Parameter constraints are obtained by minimizing the chi-squared statistic $\chi^2(p)$, which quantifies the residuals between theoretical predictions and observations weighted by the corresponding uncertainties. Assuming Gaussian errors, the likelihood is given by $\mathcal{L}(p) \propto \exp\big[-\chi^2(p)/2\big]$, so $\chi^2$ minimization is equivalent to maximum-likelihood estimation and provides the basis for Bayesian posterior inference. Popular samplers such as {\it emcee}, {\it PyMC}, and {\it Stan} implement this (log-)likelihood when modeling data with Gaussian noise.

The expansion history of the universe, encoded in the Hubble parameter $H(z)$, is often inferred using distance-ladder techniques or standard candles such as Type Ia supernovae. An alternative and complementary approach is the Cosmic Chronometer (CC) method, first proposed in \cite{Jimenez} and subsequently refined in \cite{Simon}. In contrast to distance-based probes, the CC technique provides a direct and largely model-independent estimate of $H(z)$ by exploiting the differential age evolution of passively evolving galaxies. In practice, one identifies {\it standard clocks}, typically massive early-type galaxies (ETGs), whose stellar populations evolve with minimal recent star formation.

Using the CC data listed in Table~(\ref{tab:II}) (Appendix~\ref{App:D}), we quantify the goodness of fit through the usual chi-square statistic \cite{Eisenstein1, Hu}
\ben
\chi^2_{CC}= \sum_{i=1}^{N=35} \left( \frac{H^{\rm th}(z_i,\mathbf{p})- H^{\rm obs}(z_i)}{\sigma_{\rm CC, i}} \right)^2, \label{67}
\een
where $H^{\rm th}(z_i,\mathbf{p})$ is the theoretical Hubble parameter obtained by solving Eq.~(\ref{61}) for a given parameter set $\mathbf{p}$, $H^{\rm obs}(z_i)$ is the corresponding observational value, and $\sigma_{\rm CC, i}$ denotes the $1\sigma$ uncertainty of the CC measurement reported in Table~(\ref{tab:II}).\\

The \textit{Pantheon+SHOES} compilation contains 1701 high-quality light-curve measurements corresponding to 1550 distinct Type Ia supernovae (SNe Ia), spanning the redshift range $0.00122\le z\le 2.2613$ \cite{Brout, Scolnic}. Cosmological constraints are obtained by statistically comparing the observed distance moduli inferred from the photometric data with the theoretical predictions of a given cosmological model. The distance modulus $\mu(z,\theta)$ is defined as
\ben
\mu(z,\theta)= (m-M) = 5\,\log_{10}(D_L(z))+25, 
\label{68}
\een
where $m$ and $M$ are the apparent and absolute magnitudes, respectively, and $D_L(z)$ is the luminosity distance. For a spatially flat FLRW background it is given by \cite{Weinberg}
\ben
D_L(z)=c(1+z)\int_{0}^{z}\frac{dz_1}{H(z_1)}, \label{69}
\een
with $c$ the speed of light (in km/s) and $H(z)$ in units of km/s/Mpc, so that $D_L(z)$ is obtained in Mpc. Differentiating Eq.~(\ref{69}) with respect to $z$ yields, we get the equivalent  differential equation of $D_L(z)$ as mentioned in Eq. (\ref{60}).

In our framework, Eq.~(\ref{66}) corresponds to the electromagnetic luminosity distance, i.e.
$D_L(z)\equiv D_L^{\rm EM}(z)$, consistent with the definitions introduced in Eqs.~(\ref{59}) and (\ref{60}).

For the \textit{Pantheon+SHOES} dataset, the full covariance matrix $C$ (of dimension $1701\times 1701$) accounts for both statistical and systematic uncertainties in the set of observed distance moduli. The corresponding chi-square is therefore written as
\ben
\chi^2_{SN}= \Big(\mu_{\rm th}-\mu_{\rm obs}\Big)^T C^{-1} \Big(\mu_{\rm th}-\mu_{\rm obs}\Big),
\label{70}
\een
where $\mu_{\rm th}\equiv\mu_{\rm th}(z_i,\mathbf{p})$ is computed from Eq.~(\ref{65}) (using the solutions of Eqs.~(\ref{61}), (\ref{62}), and (\ref{67}) for a given parameter set $\mathbf{p}$), $\mu_{\rm obs}\equiv\mu_{\rm obs}(z_i)$ denotes the observed moduli from Pantheon+SHOES \cite{Brout, Scolnic}, and $C^{-1}$ is the inverse covariance matrix. \\

Baryon acoustic oscillations (BAO) provide a robust standard ruler in observational cosmology and deliver stringent constraints on the expansion history of the Universe. Extracting the BAO feature from large-scale structure measurements (e.g., galaxy clustering) requires a set of cosmological distance measures to convert observed angular and redshift separations into comoving scales. In particular, the transverse comoving distance $D_M(z)$, the Hubble distance $D_H(z)$, and the volume-averaged distance $D_V(z)$ play a central role. These quantities enable precise inference of key parameters, including the present-day Hubble constant $H_0$ and the redshift-dependent dark-energy equation-of-state parameter $w(z)$, thereby providing critical tests of the $\Lambda$CDM model and its extensions such as $w$CDM.

In spectroscopic galaxy redshift surveys, the BAO signal is observed along both the radial (line-of-sight) and transverse (angular) directions, thereby encoding information about cosmic distances as well as the expansion rate of the universe. The radial component of the BAO feature, characterized by the redshift separation $\Delta z$ associated with the sound-horizon scale, enables a direct estimate of the Hubble parameter via $H(z)=c\,\Delta z/r_d$. Here,  $r_d$ denotes the comoving sound horizon at the baryon drag epoch, i.e., the maximum distance that acoustic waves in the photon-baryon plasma could travel prior to baryon decoupling. It is given by \cite{Hu,Eisenstein2}
\ben
r_d= \int^{\infty}_{z_{drag}} \frac{c_s}{H(z)} dz \label{71}
\een
where $z_{drag} (\approx 1020)$ denotes the baryon drag epoch, corresponding to the redshift at which baryons decouple from the photon–baryon fluid, and $c_s$ is the sound speed within this plasma. In the present analysis, $r_d$ is treated as a free parameter, allowing it to be directly constrained by observational data. The BAO feature also provides a measure of the Hubble distance at redshift $z$, defined as
\ben
D_H(z)=\frac{c}{H(z)}. \label{72}
\een
In the transverse direction, the BAO feature corresponds to a characteristic angular scale $\Delta\theta=r_d/D_M(z)$, i.e. the projection of the comoving sound horizon at redshift $z$. The transverse comoving distance $D_M(z)$ depends on the expansion history and can be computed from Eq.~(\ref{69}) by replacing $D_L(z)$ with $D_M(z)$ \cite{Alam, Eisenstein1, Percival, Glazebrook}.

When accounting for the cosmological dependence of $r_d$, BAO observations primarily constrain the dimensionless ratios $D_M(z)/r_d$ and $D_H(z)/r_d$. Traditionally, BAO measurements were also reported using the spherically averaged (isotropic) distance $D_V(z)$, defined as \cite{Alam, Eisenstein1, Percival, Glazebrook}
\ben
D_V(z)=\big[z\,D_M(z)^2\,D_H(z)\big]^{\frac{1}{3}}, \label{73}
\een
which combines radial and transverse information into a single observable. This compression is particularly useful at low redshift, where the sensitivity to anisotropies between line-of-sight and angular clustering is limited. In such cases, $D_V(z)$ provides a convenient proxy for the geometric constraints from BAO, facilitating comparisons between cosmological models and parameter estimation.\\

In this work, we utilize two independent BAO datasets. The first, referred to as SDSSBAO, consists of $8$ BAO measurements obtained from the \textit{Sloan Digital Sky Survey (SDSS)} \cite{Alam, Dawson}. The second dataset, denoted DESBAO, includes $9$ BAO data points derived from the \textit{Dark Energy Spectroscopic Instrument Data Release 2 (DESI DR2)} \cite{DESI}. We combine these samples into a unified BAO dataset (17 data points in total). The datasets are listed in Tables~\ref{tab:III} and \ref{tab:IV} (Appendix~\ref{App:D}). For the BAO likelihood \cite{Hogg, Hogg1}, we adopt a Gaussian chi-square of the form
\ben
\chi^2_{\rm BAO}=\sum_{i=1}^{N}\left(\frac{X^{\rm th}(z_i,\mathbf{p})-X^{\rm obs}(z_i)}{\sigma_i}\right)^2, \label{74}
\een
where $X^{\rm th}(z_i,\mathbf{p})$ denotes the theoretical prediction for the BAO observable (e.g., $D_M(z_i)/r_d$, $D_H(z_i)/r_d$, or $D_V(z_i)/r_d$), and $X^{\rm obs}(z_i)$ and $\sigma_i$ are the corresponding measurements and uncertainties. When multiple BAO observables are included, the total BAO contribution is given by
\ben
\chi^2_{\rm BAO}=\chi^2_{D_M/r_d}+\chi^2_{D_H/r_d}+\chi^2_{D_V/r_d}.
\label{74a}
\een\\

We next describe the gravitational-wave (GW) datasets. GW astronomy provides an independent probe of the cosmic expansion history through standard sirens. Observations by the LIGO, Virgo, and KAGRA collaborations \cite{GWOSC} have resulted in successive releases of the Gravitational-Wave Transient Catalogs (GWTC), which provide posterior distributions for source parameters---including the luminosity distance---that can be used for cosmological inference. In this work, we use luminosity-distance measurements from GWTC-2.1 \cite{GWTC2.1}, GWTC-3 \cite{GWTC3}, and GWTC-4 \cite{GWTC4, Abac, Abac2}, which span multiple observing runs and improve both the statistical power and redshift coverage of standard-siren analyses.

The GWTC-2.1 catalog \cite{GWTC2.1} is an extended and refined version of GWTC-2 \cite{GWTC2}, based on a comprehensive reanalysis of events detected during the first half of the third observing run (O3a) of Advanced LIGO and Advanced Virgo. It includes confidently identified binary black-hole (BBH) mergers and previously reported sources such as binary neutron-star (BNS) systems. A key feature of GWTC-2.1 is its \emph{deep} reanalysis, which employs improved data-quality vetting, mitigation of instrumental artifacts (glitches), and multiple matched-filter pipelines. By adopting a more permissive false-alarm-rate (FAR) threshold, the catalog reports additional high-confidence candidates and increases the number of significant O3a detections to 44. These methodological improvements yield more precise and reliable posterior distributions for source parameters, particularly the luminosity distance, which is central to standard-siren cosmology. Consequently, GWTC-2.1 provides a statistically rich dataset for intermediate-redshift standard-siren analyses.

The GWTC-3 catalog \cite{GWTC3} provides a comprehensive compilation of detections from the full third observing run (O3), including both O3a and O3b. It substantially increases the sample of compact-binary coalescences, reporting 90 confident gravitational-wave events. The catalog includes a diverse population of sources, including binary black hole (BBH) mergers, binary neutron-star (BNS) mergers, and the first confirmed neutron-star-black-hole (NSBH) mergers (e.g., GW200105 and GW200115). The enlarged event sample strengthens population studies and enables more informative hierarchical analyses. Moreover, GWTC-3 extends the redshift reach of GW observations, improving constraints on the luminosity-distance--redshift relation. These advances are driven by improved detector sensitivity, refined waveform models, and (for a subset of events) the inclusion of KAGRA, which together enhance sky localization and help reduce degeneracies between luminosity distance and binary inclination. In a cosmological context, GWTC-3 provides an important foundation for \emph{dark siren} analyses, in which GW luminosity distances are statistically associated with galaxy catalogs to infer cosmological parameters such as $H_0$.

The GWTC-4 catalog \cite{GWTC4, Abac, Abac2} reports detections from the first half of the fourth observing run (O4a), using data from Advanced LIGO, Advanced Virgo, and KAGRA operating at improved (``A+'') sensitivities. The rapid increase in the number of reported candidates substantially enlarges the event sample and enhances the statistical power of GW datasets. The inclusion of a multi-detector network improves sky localization and helps reduce degeneracies between luminosity distance and orbital inclination, enabling more precise parameter estimation. Increased sensitivity also extends the reach to higher-redshift compact-binary mergers, thereby probing the luminosity-distance-redshift relation in previously inaccessible regimes. In addition, the catalog includes several high-mass BBH events that serve as useful laboratories for tests of gravity and cosmology. Owing to the larger number of high-confidence detections, statistical uncertainties on luminosity-distance measurements are reduced, making GWTC-4 particularly valuable for precision-cosmology applications. Overall, GWTC-4 provides a comprehensive dataset for constraining the expansion history of the universe, studying merger-rate evolution, and searching for potential deviations from standard cosmology, including modified GW propagation.\\

To incorporate the gravitational-wave (GW) luminosity-distance measurements from the GWTC-2.1 \cite{GWTC2.1}, GWTC-3 \cite{GWTC3}, and GWTC-4 \cite{GWTC4, Abac, Abac2} catalogs into our joint likelihood analysis, we construct a standard chi-square estimator for the GW sector \cite{Zhang1}. The LVK collaboration \cite{GWOSC} reports the GW luminosity distance $D_L^{\rm GW}(z)$ with asymmetric $90\%$ credible intervals, quoted as upper and lower uncertainties $\sigma^+_{\rm GW}$ and $\sigma^-_{\rm GW}$. These asymmetries reflect the non-Gaussian nature of the distance posterior and are driven mainly by the degeneracy between luminosity distance and binary inclination, together with detector noise properties. Following Refs.~\cite{Zhang1, Cai}, we approximate the asymmetric uncertainties by an effective symmetric standard deviation,
\ben
\sigma_{\rm GW}=\frac{\sigma^+_{\rm GW}+\sigma^-_{\rm GW}}{3.29}, \label{75}
\een
where the factor $3.29\simeq 2\times 1.645$ converts a two-sided $90\%$ credible interval to an equivalent Gaussian $1\sigma$ uncertainty \cite{GWTC2.1, GWTC3, GWTC4}. The GW contribution to the likelihood is then written as
\ben
\chi^2_{\rm GW}=\sum_{i=1}^{N}\left(\frac{D^{\rm GW,th}_L(z_i,\mathbf{p})-D^{\rm GW,obs}_L(z_i)}{\sigma_{{\rm GW},i}}\right)^2, \label{76}
\een
where $D^{\rm GW,th}_L(z_i,\mathbf{p})$ is the theoretical GW luminosity distance obtained by solving Eq.~(\ref{60}), and $D^{\rm GW,obs}_L(z_i)$ denotes the measured values from the GWTC catalogs. In our analysis, we consider three GW data combinations, namely GWTC (2.1+3), GWTC-4, and GWTC (2.1+3+4). For dimensional consistency, all reported $D_L^{\rm GW}$ values are converted from Gpc to Mpc.
\\

We employed Bayesian inference using the No-U-Turn Sampler (NUTS) \cite{Gelman} to constrain the cosmological parameters $\mathbf{p}$ from the OHD dataset. In this framework, Markov chain Monte Carlo (MCMC) sampling \cite{Lewis} is used to generate posterior distributions for the model parameters and to quantify the associated uncertainties. Point estimates are reported using the posterior mean values obtained from the MCMC chains. To assess the overall goodness of fit, we also compute the chi-square statistic, which measures the agreement between theoretical predictions and observational data. The resulting marginalized posteriors are visualized through 1D distributions and 2D contour plots, showing the $1\sigma$ (68\%) and $2\sigma$ (95\%) credible regions \cite{Lin, De} and highlighting parameter correlations in the multidimensional space.

\section{Results and Discussion} \label{sec.VI}
\label{Sec:VI}
\begin{figure}
\centering
\includegraphics[width=7.9cm, height=7.8cm]{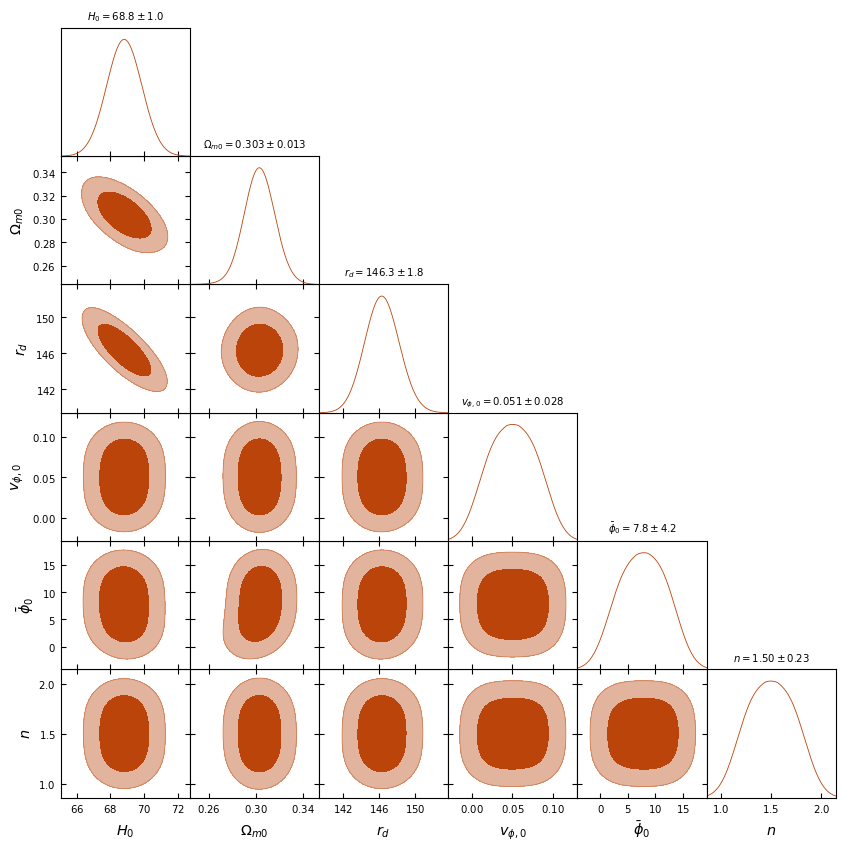}
 \caption{Posterior constraints on the k-essence model parameters from the combined CC+BAO dataset.}\label{Fig2} 
\end{figure}

\begin{figure}
\centering
\includegraphics[width=7.9cm, height=7.8cm]{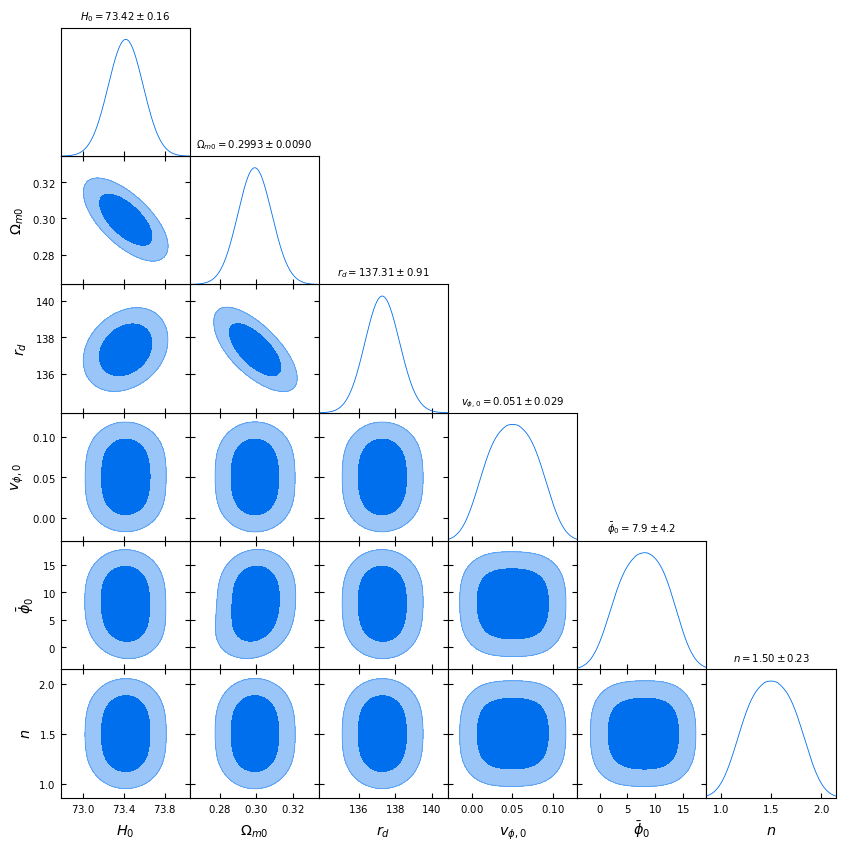}
 \caption{Posterior constraints on the k-essence model parameters from the CC+Pantheon+BAO data combination.}\label{Fig3} 
\end{figure}

In our Bayesian analysis, we set the initial conditions and assign minimally informative uniform priors to the main model parameters $p=\{H_0,\Omega_{m,0},v_{\phi,0},\bar{\phi}_0,n,r_d\}$ (Table~\ref{tab:II}), where $\mathcal{U}(a,b)$ denotes a uniform distribution on $(a,b)$. We choose physically reasonable prior ranges to limit prior-driven bias and to let the data dominate the constraints. To keep the sound speed well defined, $0<c_s^2\leq 1$, and to preserve causality in the interacting k-essence model, we restrict the kinetic index to $n\sim\mathcal{U}(1.1,1.9)$ (Table~\ref{tab:II}). Figures~\ref{Fig2} and \ref{Fig3} present the 1D marginalized posteriors and 2D contour plots, and we quote the $1\sigma$ and $2\sigma$ credible intervals for the cosmological and model-specific parameters. Figure~\ref{Fig2} shows the constraints from the combined CC+BAO analysis \cite{Jimenez, zhang, Jimenez1, Simon, Stern, Moresco, Abbott2, Metin, Eric, Julian, Betoule, Alam, Eisenstein1, Percival, Glazebrook, Hu, Eisenstein2, Dawson, DESI}. As shown in Fig.~\ref{Fig3}, adding Pantheon+ \cite{Brout} further breaks remaining background degeneracies, giving our tightest and most robust constraints. For completeness, we also analyze the GWTC (2.1+3+4) dataset alone \cite{GWTC2.1, GWTC3, GWTC4}, with results shown in Fig.~\ref{Fig7}.

CMB-based measurements of the expansion rate, especially those from the \,Planck Collaboration, tend to favor a lower Hubble constant within a flat $\Lambda$CDM model. The 2018 \,Planck analysis finds $H_0=67.36\pm 0.54\,\mathrm{km\,s^{-1}\,Mpc^{-1}}$ \cite{Aghanim, Planck2}, and the 2020 legacy release reports a consistent value of $H_0\approx 67.4\pm 0.5\,\mathrm{km\,s^{-1}\,Mpc^{-1}}$, supporting the stability of the CMB-based inference \cite{Planck3}. By contrast, late-time, local measurements such as those from the SH0ES Collaboration \cite{Riess} favor a higher value, $H_0\approx 73.2\pm 1.3\,\mathrm{km\,s^{-1}\,Mpc^{-1}}$. This long-standing mismatch between early- and late-universe estimates is commonly known as the ``Hubble tension.''

We constrain the cosmological parameters $\{H_0,\Omega_{m,0},r_d\}$ together with the model parameters $\{v_{\phi,0},\bar{\phi}_0,n\}$ using Bayesian MCMC inference. As expected, combining complementary datasets sharpens the constraints and makes the results more stable. Using CC+BAO (Fig.~\ref{Fig2}) \cite{Jimenez, zhang, Jimenez1, Simon, Stern, Moresco, Abbott2, Metin, Eric, Julian, Betoule, Alam, Eisenstein1, Percival, Glazebrook, Hu, Eisenstein2, Dawson, DESI}, we find $H_0=68.8\pm 1.0\,\mathrm{km\,s^{-1}\,Mpc^{-1}}$, broadly consistent with early-universe background expansion. Using only the GWTC (2.1+3+4) dataset (Fig.~\ref{Fig7}) \cite{GWTC2.1, GWTC3, GWTC4}, we obtain $H_0=68.1\pm 1.7\,\mathrm{km\,s^{-1}\,Mpc^{-1}}$. Our tightest bound comes from the full CC+Pantheon+BAO combination (Fig.~\ref{Fig3}) \cite{Jimenez, zhang, Jimenez1, Simon, Stern, Moresco, Abbott2, Metin, Eric, Julian, Betoule, Alam, Eisenstein1, Percival, Glazebrook, Hu, Eisenstein2, Dawson, DESI, Brout}, which gives $H_0=73.42\pm 0.16\,\mathrm{km\,s^{-1}\,Mpc^{-1}}$. This higher late-time value is close to local distance-ladder measurements \cite{Riess} but remains in strong tension with the Planck 2018 CMB inference \cite{Planck3}, $H_0=67.4\pm 0.5\,\mathrm{km\,s^{-1}\,Mpc^{-1}}$, reinforcing the persistence of the Hubble tension.

These same analyses also constrain the sound horizon at the drag epoch. From Fig.~\ref{Fig2} (CC+BAO), we obtain $r_d=146.3\pm 1.8\,\mathrm{Mpc}$, while the full CC+Pantheon+BAO combination (Fig.~\ref{Fig3}) favors a smaller value, $r_d=137.31\pm 0.91\,\mathrm{Mpc}$ \cite{Brout}. The inclusion of Pantheon+ thus shifts $r_d$ downward and, at the same time, raises the preferred Hubble constant to $H_0=73.42\pm 0.16\,\mathrm{km\,s^{-1}\,Mpc^{-1}}$. This anticorrelated behavior is a manifestation of the well-known $H_0$--$r_d$ degeneracy in BAO distance measurements \cite{Eisenstein1, Alam, Dawson}. As a result, the full dataset can accommodate an $H_0$ value close to local distance-ladder estimates \cite{Riess}; however, this does not by itself resolve the Hubble tension, since it remains inconsistent with the CMB-inferred value from Planck \cite{Planck3}. The fact that different dataset combinations prefer different $(H_0,r_d)$ pairs indicates that the tension is not alleviated uniformly across probes. In our framework, this shift is enabled in part because $r_d$ is treated as a free parameter, allowing the fit to move along the BAO degeneracy direction. Therefore, rather than providing a universal solution to the Hubble tension, the k-essence model exhibits sufficient phenomenological flexibility to fit a higher late-time expansion rate (as required by supernova luminosity distances) while remaining compatible with BAO angular-scale constraints. More generally, k-essence models can exhibit attractor-like behavior, making their late-time evolution less sensitive to the choice of initial conditions, and they were originally introduced as a mechanism to drive cosmic acceleration without the severe fine-tuning required by a pure cosmological constant \cite{Picon2, Picon, Vikman}.

Using the combined CC+Pantheon+BAO dataset (Fig.~\ref{Fig3} and Table~\ref{tab:II}), which yields our tightest constraints, we find $\bar{\phi}_0 = 7.9\pm 4.2\,\mathrm{eV}$ for the present-day background field, $v_{\phi,0}=0.051\pm 0.029\,(\mathrm{eV})^2$ for its current velocity, and $n=1.50\pm 0.23$ for the kinetic index. The preferred value $n\simeq 1.5$ falls within the physically allowed range $1<n<2$, which guarantees a well-defined sound speed, $0<c_s^2\leq 1$. This indicates a departure from canonical scalar-field dynamics and highlights the role of the non-linear kinetic structure in the Lagrangian $\mathcal{L}=X^n - V(\phi)$. In particular, $v_{\phi,0}$ governs the field's present-day evolution, allowing the model to mimic an effective cosmological constant while still retaining dynamical freedom. The posterior correlation matrices for the parameter set $p=\{H_0,\,\Omega_{m,0},\,r_d,\,v_{\phi,0},\,\bar{\phi}_0,\,n\}$ are shown in Figs.~\ref{Fig8} and \ref{Fig9} (Appendix~\ref{App:D}) for the CC+BAO and CC+Pantheon+BAO analyses, respectively.

\begin{table*}[t]
    \centering
    \renewcommand{\arraystretch}{1.3}
    \caption{Posterior constraints on cosmological and k-essence parameters for the different dataset combinations considered.}
     \label{tab:II}
     \begin{tabular}{c c c c c }
        \hline\hline
        \textbf{Parameters} & \textbf{Prior} & \multicolumn{3}{c|}{\textbf{Datasets}} \\ \cline{3-5} 
 & & \textbf{CC + BAO } & \textbf{CC + Pantheon + BAO } & \textbf{GWTC (2.1+3+4) only} \\ \hline
$H_0$ [Km/s/Mpc] & $\mathcal{U}(60, 80)$ & $68.8 \pm 1.0$ & $73.42 \pm 0.16$ & $68.1 \pm 1.7$ \\ 
$\Omega_{m,0}$ & $\mathcal{U}(0.2, 0.4)$ & $0.303 \pm 0.013$ & $0.2993 \pm 0.009$ & $0.302 \pm 0.056$ \\ 
$r_d$ [Mpc] & $\mathcal{U}(130, 160)$ & $146.3 \pm 1.8$ & $137.31 \pm 0.91$ & --\\ \hline
$v_{\phi,0}$ [eV$^2$] & $\mathcal{U}(0.01, 0.1)$ & $0.051 \pm 0.028$ & $0.051 \pm 0.029$ & $0.050 \pm 0.029$ \\ 
$\bar{\phi}_0$ [eV] & $\mathcal{U}(0.1, 15.0)$ & $7.8 \pm 4.2$ & $7.9 \pm 4.2$ & $7.5 \pm 4.3$ \\ 
$n$ & $\mathcal{U}(1.1, 1.9)$ & $1.5 \pm 0.23$ & $1.5 \pm 0.23$ & $1.5 \pm 0.23$\\ 
    \hline\hline
    \end{tabular}
\end{table*}

\begin{figure}[H]
\centering
\includegraphics[width=7.8cm, height=4.7cm]{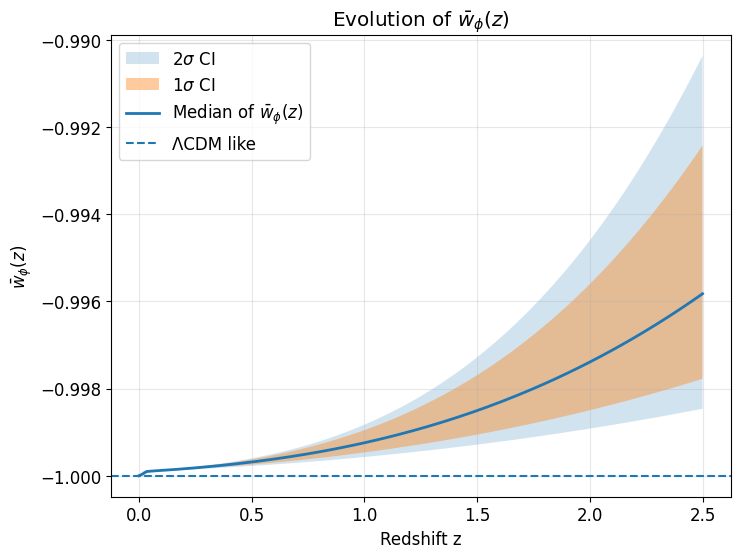}
 \caption{Evolution of the background scalar-field equation-of-state parameter $\bar{w}_{\phi}(z)$ as a function of redshift, reconstructed using the best-fit parameters from the combined CC+BAO+Pantheon dataset. The solid curve represents the posterior median, while the shaded regions denote the $1\sigma$ and $2\sigma$ credible intervals. The dashed horizontal line indicates the $\Lambda$CDM value, $w=-1$}\label{Fig4} 
\end{figure}

Using the best-fit parameters from the combined CC+BAO+Pantheon analysis, we reconstruct the dark-energy dynamics. Figure~\ref{Fig4} shows the reconstructed background scalar-field equation-of-state (EoS) $\bar{w}_\phi(z)$. At $z=0$ we obtain $\bar{w}_{\phi,0}\simeq -1$, i.e. the k-essence background is effectively indistinguishable from a cosmological constant at late times. However, the apparently ``ultra-tight'' constraint on $\bar{w}_{\phi,0}$ should not be read as an observational determination at the $10^{-7}$ level: $\bar{w}_{\phi,0}$ is not sampled directly in the MCMC, but is a derived quantity fixed by the sampled parameters (in our parameterization, mainly $v_{\phi,0}$ and $n$ through the kinetic-to-total energy-density ratio). In the region allowed by the fit the kinetic contribution at $z=0$ is forced to be strongly subdominant, so the reconstruction is driven toward the potential-dominated limit $\bar{w}_{\phi,0}\to -1$. Consistent with this interpretation, a prior-only propagation gives $\sigma_{\rm prior}\sim 9.03\times 10^{-8}$, very close to the posterior value $\sigma_{\rm posterior}\sim 7.73\times 10^{-8}$, showing that the numerical precision largely comes from the adopted parameterization and prior ranges rather than independent information in the data. At higher redshift the credible bands widen and small deviations from $w_{\rm de}=-1$ appear, reflecting the model's mild dynamics at earlier epochs and its approach to $\Lambda$-like behavior toward $z=0$.\\

\begin{figure}[H]
\centering
\includegraphics[width=7.8cm, height=4.7cm]{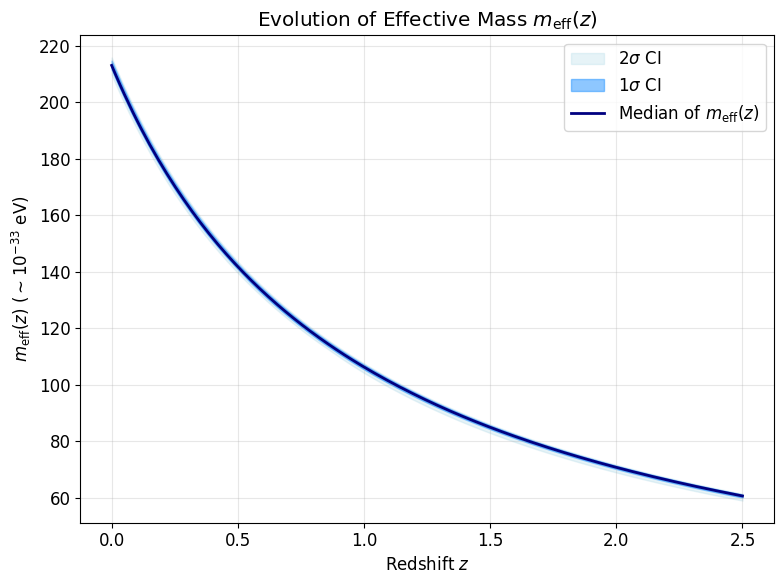}
\caption{Redshift evolution of the effective mass term $m_{\rm eff}(z)$ (of order $10^{-33}\,\mathrm{eV}$) reconstructed using the best-fit parameters from the CC+Pantheon++BAO analysis. The shaded bands show the $1\sigma$ and $2\sigma$ credible intervals.}\label{Fig5}
\end{figure}

A central outcome of this work is the reconstructed redshift dependence of the effective mass scale $m_{\rm eff}(z)$ shown in Fig.~\ref{Fig5}. From the modified gravitational-wave equation~\eqref{50}, the quantity $m_{\rm eff}^2$ carries the same dimensions as $k^2$, namely inverse length squared (equivalently inverse time squared). Consequently, $m_{\rm eff}$ has units of inverse length or inverse time and therefore defines an energy scale, which can be interpreted as a mass scale in natural units $(c=\hbar=1)$ (see Appendix~\ref{App:C}). Since $m_{\rm eff}^2$ is determined by background scalar-field quantities, specifically, Eq.~\eqref{48} implies $m_{\rm eff}^2\propto(\bar{\rho}_{\phi}-\bar{P}_{\phi})$, its characteristic magnitude is set by the cosmological background scale, and is therefore expected to be of order $H^2$ (see Eq.~\eqref{58}). Using the conversion $1\,\mathrm{km\,s^{-1}\,Mpc^{-1}}\simeq 2.13\times 10^{-42}\,\mathrm{GeV}\simeq 2.13\times 10^{-33}\,\mathrm{eV}$, we find that the implied scale is ultra-light, $m_{\rm eff}\sim\mathcal{O}(10^{-33})\,\mathrm{eV}$, placing it firmly in the cosmological ultra-light regime~\cite{Unit}.

For the CC+Pantheon++BAO best-fit combination, we obtain a present-day effective mass $m_{\rm eff,0} \approx 212.88 \pm 1.72\,\mathrm{km\,s^{-1}\,Mpc^{-1}}$. Interpreting this as a frequency (or inverse-length) scale and converting units as discussed above, it corresponds to an ultra-light mass scale of order $10^{-33}\,\mathrm{eV}$.

It is also important to clarify how $m_{\rm eff}$ enters our inference. In our MCMC analysis, $m_{\rm eff}$ is not sampled as an independent parameter (and neither is the scalar-field equation-of-state $\bar{w}_\phi$). Instead, $m_{\rm eff}(z)$ is a derived quantity fixed by the underlying k-essence background once the primary parameters are specified. Accordingly, the combined datasets constrain the fundamental background parameters
$\{H_0,\,\Omega_{m,0},\,r_d,\,v_{\phi,0},\,\bar{\phi}_0,\,n\}$.
Given posterior samples of these parameters, we then reconstruct the background evolution and compute the derived functions $\bar{w}_\phi(z)$ and $m_{\rm eff}(z)$. Therefore, the redshift dependence of $m_{\rm eff}(z)$, and its impact on gravitational-wave propagation, is not fitted directly to the data, but emerges as a prediction of the k-essence model conditioned on the observationally constrained background parameters.\\

Figure~\ref{Fig5} shows the evolution of $m_{\rm eff}$ (in units of $\mathcal{O}(10^{-33})\,\mathrm{eV}$). Its redshift dependence follows from the k-essence background via $(\bar{\rho}_{\phi}-\bar{P}_{\phi})=(1-\bar{w}_{\phi})\bar{\rho}_{\phi}$ in Eq.~\eqref{48}, together with the overall $a^2$ scaling. Because the k-essence field sources late-time acceleration and approaches $\bar{w}_{\phi,0}\simeq -1$ at low redshift, $m_{\rm eff}$ is maximal today and decreases monotonically with increasing redshift, as seen in Fig.~\ref{Fig5}. The reconstructed $m_{\rm eff}(z)$ thus encapsulates the redshift-dependent effective-mass term induced by the k-essence background. This term alters the gravitational-wave phase during propagation, producing a frequency-dependent effect, while the principal part of the propagation equation still implies a luminal characteristic speed.

\begin{figure}
\centering
\includegraphics[width=7.8cm, height=4.7cm]{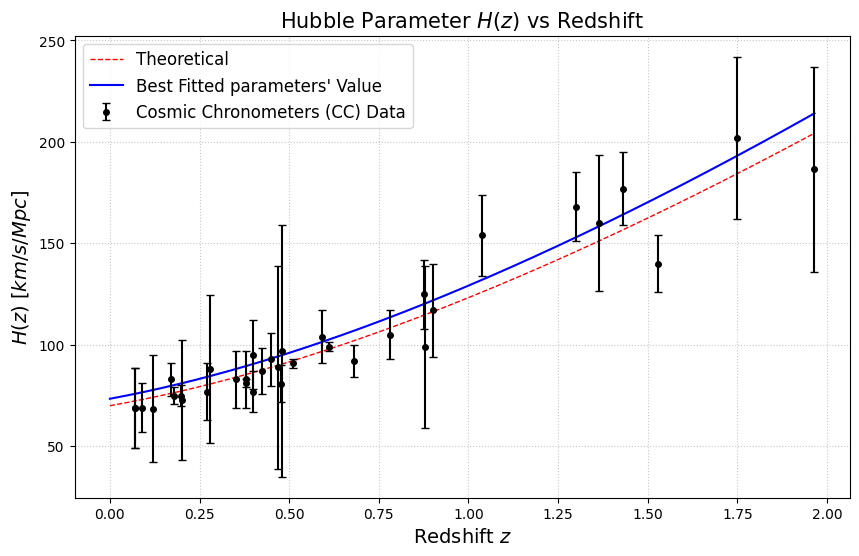}
\caption{All available CC data points are listed in Table~(\ref{tab:V}) in Appendix~\ref{App:D}. The dashed red curve shows the theoretical Hubble parameter $H(z)$ computed using the fiducial values $\Omega_{\rm de,0}=0.7$, $\Omega_{m,0}=0.3$, and $H_0=70.0\,\mathrm{km\,s^{-1}\,Mpc^{-1}}$. The solid blue curve shows $H(z)$ computed using the best-fit parameters from our CC+Pantheon+BAO analysis.}\label{Fig6}
\end{figure}

Finally, Fig.~\ref{Fig6} compares the reconstructed Hubble parameter $H(z)$ with cosmic-chronometer measurements. The reconstructed prediction (solid curve) agrees well with the data across the full redshift range, showing that the model reproduces the observed expansion history while still permitting non-trivial, testable modifications in the gravitational-wave sector.\\

We also performed a Bayesian analysis of the $\Lambda$CDM model using the CC+BAO, CC+BAO+Pantheon, and GWTC (2.1+3+4)-only datasets. The corresponding two-dimensional corner plots are shown in Figs.~\ref{Fig10} and \ref{Fig11} (Appendix~\ref{App:E}). The inferred parameter values, including the Hubble constant $H_0$, the present-day dark-energy density parameter $\Omega_{\rm de,0}$, and the sound horizon $r_d$, are summarized in Table~\ref{tab:VIII} (Appendix~\ref{App:E}). Among these combinations, CC+BAO+Pantheon provides the tightest constraints and yields the overall best-fit $\Lambda$CDM parameter values.\\

\begin{table*}[t]
    \centering
    \renewcommand{\arraystretch}{1.3}
    \caption{Comparison of statistical criteria for k-essence and $\Lambda$CDM }
    \label{tab:III}
    \begin{tabular}{c c c c c c}
        \hline\hline
        \textbf{Model} & \textbf{$\chi^2$} & \textbf{$\chi^2_{\rm red}$} & \textbf{AIC} & \textbf{BIC} & \textbf{DIC} \\ \hline 
        $\Lambda$CDM (CC+BAO+Pantheon) & $1840.20$ & $1.0462$ & $1846.20$ & $1862.62$ & $1846.19$\\ \hline
        k-essence (CC+BAO+Pantheon) & $1836.23$ & $1.0457$ & $1848.23$ & $1881.077$ & $1842.36$\\ \hline\hline
    \end{tabular}
\end{table*}

To compare how effectively the k-essence and $\Lambda$CDM models fit the combined observational dataset (CC+BAO+Pantheon), we consider several standard goodness-of-fit and model-selection statistics. We compute the total chi-squared, $\chi^2$, and the reduced chi-squared, $\chi^2_{\rm red}=\frac{\chi^2}{\nu},\quad \nu=N-k,$ where $\nu$ is the number of degrees of freedom, $N$ is the total number of data points in the combined dataset, and $k$ is the number of free parameters in the model. In addition, we evaluate three information criteria:
\begin{itemize}
    \item \textbf{AIC} (Akaike information criterion): $\mathrm{AIC}=\chi^2+2k$.
    \item \textbf{BIC} (Bayesian information criterion): $\mathrm{BIC}=\chi^2+k\ln N$.
    \item \textbf{DIC} (deviance information criterion): $\mathrm{DIC}=\overline{D}(\theta)+p_D$, where $\overline{D}(\theta)$ is the posterior mean deviance and $p_D$ is the effective number of parameters.
\end{itemize}

We therefore compare five metrics ($\chi^2$, $\chi^2_{\rm red}$, AIC, BIC, and DIC), summarized in Table~\ref{tab:III}. In our case, the number of data points is the same for both models ($N=1762$), while the parameter counts are $k=6$ for the k-essence model and $k=3$ for $\Lambda$CDM. These criteria quantify goodness of fit while penalizing overly complex models, thereby helping to assess potential overfitting.

Table~\ref{tab:III} compares the statistical performance of the standard $\Lambda$CDM model and our k-essence model using the combined CC+BAO+Pantheon dataset. The k-essence model has a slightly smaller $\chi^2$ than $\Lambda$CDM, suggesting a marginally better fit in an absolute sense. However, the difference in $\chi^2$ is very small, so both models describe the data nearly equally well. This is also reflected in the reduced chi-squared values, $\chi^2_{\rm red}$, which are both close to unity. Overall, based on these goodness-of-fit measures, the k-essence model performs comparably to $\Lambda$CDM.\\ 

Information criteria provide a complementary comparison by explicitly accounting for the different parameter counts of the two frameworks. Because the k-essence model introduces additional free parameters, it incurs a larger complexity penalty. For the AIC we find $\mathrm{AIC}=1848.23$ (k-essence) and $\mathrm{AIC}=1846.20$ ($\Lambda$CDM) in Table~\ref{tab:III}, yielding $\Delta\mathrm{AIC}\equiv(\mathrm{AIC_{k\text{-}essence}}-\mathrm{AIC_{\Lambda\mathrm{CDM}}})=+2.03$. Under the usual rule of thumb, $|\Delta\mathrm{AIC}|\lesssim2$ suggests little to no practical distinction, while larger differences increasingly favor the model with the smaller AIC \cite{Burnham, Kass}. Thus, AIC gives a slight preference to $\Lambda$CDM, but the difference is marginal and sits close to the conventional ``comparable support'' threshold. By contrast, the BIC values are $\mathrm{BIC}=1881.077$ (k-essence) and $\mathrm{BIC}=1862.62$ ($\Lambda$CDM), corresponding to $\Delta\mathrm{BIC}=+18.457$. Since $\Delta\mathrm{BIC}>10$ is typically interpreted as strong evidence for the model with the lower BIC \cite{Burnham, Kass}, the BIC strongly favors the simpler $\Lambda$CDM framework. This is consistent with how BIC behaves for large samples: it penalizes extra parameters more severely, so any improvement in the best-fit likelihood (equivalently, the reduction in $\chi^2$) for k-essence is not large enough to compensate for its increased flexibility.\\

On the other hand, the DIC values are $\mathrm{DIC}=1842.36$ for k-essence and $\mathrm{DIC}=1846.19$ for $\Lambda$CDM (Table~\ref{tab:III}), giving $\Delta\mathrm{DIC}\equiv(\mathrm{DIC_{k\text{-}essence}}-\mathrm{DIC_{\Lambda\mathrm{CDM}}})=-3.83$. Since $2<|\Delta\mathrm{DIC}|\leq7$, this corresponds to moderate (but not decisive) support for the model with the lower DIC \cite{Best}, i.e., k-essence. The DIC is particularly useful here because its complexity penalty is set by the \emph{effective} number of parameters inferred from the posterior, rather than by the nominal parameter count alone. In this sense, DIC suggests that the modest improvement in fit achieved by k-essence remains competitive once its effective model complexity is taken into account. Taken together, the criteria give a mixed picture: the AIC difference is small (comparable support), the BIC strongly favors the simpler $\Lambda$CDM model, while the slightly lower $\chi^2$ and DIC indicate that k-essence can provide a marginally improved description of the combined CC+BAO+Pantheon dataset. Overall, neither framework is unambiguously preferred; instead, the conclusion depends on how one trades off improved fit against increased model complexity.\\

\begin{table*}[t]
    \centering
    \renewcommand{\arraystretch}{1.3}
    \caption{ Posterior predictive summaries for joint datasets}
     \label{tab:IV}
     \begin{tabular}{c c c c c c}
        \hline\hline
        \textbf{Model} & \textbf{Observable} &  \textbf{N} & \textbf{$\chi^2_{\rm med}$} & \textbf{$\chi^2_{\rm med}/N$} & \textbf{$p_B$}\\ \hline
\multirow{4}{*}{k-essence (CC+BAO)} 
& $H(z)$ & $35$ & $18.039$ & $0.515$ & $0.988$ \\ 
& $D_M/r_d$ & $11$ & $11.004$ & $1.000$ & $0.414$ \\
& $D_H/r_d$ & $11$ & $15.630$ & $1.421$ & $0.156$ \\
& $D_V/r_d$ & $4$ & $2.005$ & $0.501$ & $0.714$ \\ \hline
\multirow{4}{*}{k-essence (CC+BAO+Pantheon)} 
& $H(z)$ & $35$ & $46.732$ & $1.335$ & $0.114$ \\
& $\mu$ & $1701$ & $1761.521$ & $1.036$ & $0.156$ \\
& $D_M/r_d$ & $11$ & $11.249$ & $1.023$ & $0.391$ \\
& $D_H/r_d$ & $11$ & $15.995$ & $1.454$ & $0.138$ \\
& $D_V/r_d$ & $4$ & $2.360$ & $0.590$ & $0.644$ \\ \hline 
\multirow{4}{*}{$\Lambda$CDM (CC+BAO)} 
& $H(z)$ & $35$ & $17.852$ & $0.510$ & $0.990$ \\
& $D_M/r_d$ & $11$ & $11.579$ & $1.053$ & $0.382$ \\
& $D_H/r_d$ & $11$ & $16.299$ & $1.482$ & $0.126$ \\
& $D_V/r_d$ & $4$ & $1.799$ & $0.450$ & $0.746$ \\ \hline
\multirow{4}{*}{$\Lambda$CDM (CC+BAO+Pantheon)} 
& $H(z)$ & $35$ & $46.919$ & $1.341$ & $0.101$ \\
& $\mu$ & $1701$ & $1765.283$ & $1.038$ & $0.127$ \\
& $D_M/r_d$ & $11$ & $11.259$ & $1.024$ & $0.402$ \\
& $D_H/r_d$ & $11$ & $15.925$ & $1.448$ & $0.131$ \\
& $D_V/r_d$ & $4$ & $2.325$ & $0.581$ & $0.646$ \\ 
    \hline\hline
    \end{tabular}
\end{table*} 

To further assess the agreement between the fitted models and the observational data, we perform posterior predictive checks using the posterior distribution of the model parameters. Posterior predictive checking offers a Bayesian framework for evaluating model adequacy by comparing the observed data with replicated datasets generated from the posterior predictive distribution \cite{Gelman}. In the present analysis, we take the $\chi^2$ statistic as the discrepancy measure and summarize its posterior distribution by the median $\chi^2_{\rm med}$, the normalized median $\chi^2_{\rm med}/N$, and the posterior predictive probability $p_B$.

Here, $\chi^2_{\rm med}$ is the median of the posterior distribution of the $\chi^2$ discrepancy, and $\chi^2_{\rm med}/N$ is this median normalized by the number of data points $N$ in the corresponding observational subset. Values of $\chi^2_{\rm med}/N$ (Table~\ref{tab:IV}) close to unity indicate that the typical discrepancy is of the expected magnitude. This quantity should not be confused with the conventional reduced chi-squared, $\chi^2_{\rm red}=\chi^2/(N-k)$, reported in Table~\ref{tab:III}, where $k$ denotes the number of fitted parameters. The posterior predictive probability $p_B$ measures how extreme the observed discrepancy is relative to discrepancies computed from replicated datasets generated under the fitted model. Values of $p_B$ close to 0 or 1 suggest a potential lack of fit, whereas values away from these boundaries indicate no strong posterior-predictive discrepancy \cite{Gelman}.

Table~\ref{tab:IV} summarizes the posterior predictive results for the k-essence and $\Lambda$CDM models. For the CC+BAO joint analysis, the normalized posterior predictive discrepancies are very similar for the two models across all observables. For example, for $H(z)$ we find $\chi^2_{\rm med}/N=0.515$ for k-essence and $0.510$ for $\Lambda$CDM, and the corresponding values for $D_M/r_d$, $D_H/r_d$, and $D_V/r_d$ are likewise comparable. The associated $p_B$ values follow the same pattern, indicating no meaningful difference in posterior predictive performance between the two models for this dataset combination.

More importantly, for the combined CC+BAO+Pantheon dataset, which provides the most comprehensive joint constraint used in our analysis, the posterior predictive performance of the two models remains very similar. For $H(z)$, we obtain $\chi^2_{\rm med}/N=1.335$ for k-essence and $1.341$ for $\Lambda$CDM. For the Pantheon distance modulus, $\mu$, the corresponding values are 1.036 and 1.038, and for $D_M/r_d$ they are 1.023 and 1.024. The remaining BAO observables show the same level of agreement, with $\chi^2_{\rm med}/N=1.454$ and 1.448 for $D_H/r_d$, and 0.590 and 0.581 for $D_V/r_d$, for k-essence and $\Lambda$CDM, respectively. The associated $p_B$ values are likewise broadly consistent across all observational sectors, with no evidence for an extreme posterior predictive discrepancy in either model.

This close agreement for the CC+BAO+Pantheon combination is particularly relevant because it is the most informative joint dataset in our analysis and provides the parameter constraints used to compare the two models. The posterior predictive results therefore indicate that, despite differences in parameterization and in the number of free parameters, both k-essence and $\Lambda$CDM reproduce the individual observational components of this dataset with almost the same predictive performance.

Overall, the posterior predictive analysis indicates that the k-essence model may offer a statistically acceptable description of the data, with predictive performance comparable to that of the standard $\Lambda$CDM model. This conclusion is most clearly seen for the CC+BAO+Pantheon combination, which constitutes our primary dataset for parameter estimation, and it is consistent with the global statistical comparison reported in Table~\ref{tab:IV}.

\begin{figure}[H]
\centering
\includegraphics[width=7.9cm, height=7.8cm]{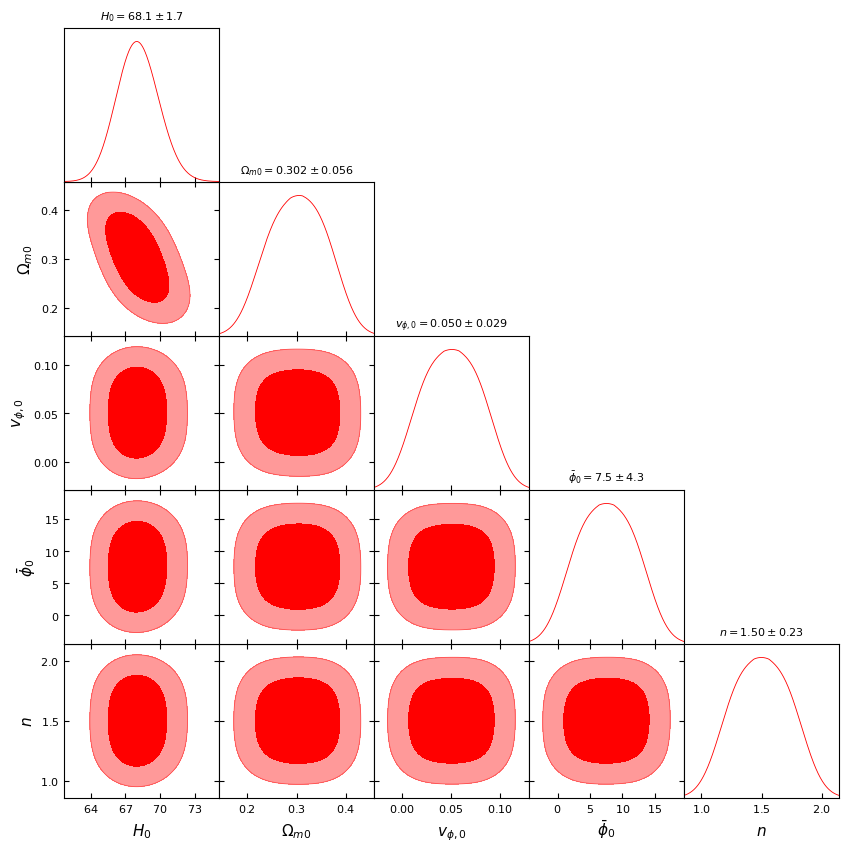}
 \caption{Fitting the k-essence model parameters using only the GWTC (2.1+3+4) dataset.}\label{Fig7} 
\end{figure}

We also perform an observational data analysis for both the k-essence and $\Lambda$CDM models using the GWTC(2.1+3+4)-only dataset. Unlike conventional electromagnetic probes, such as Cosmic Chronometers (CC), Baryon Acoustic Oscillations (BAO), and Type Ia Supernovae (Pantheon+), for which the source redshift is obtained from electromagnetic observations, gravitational-wave (GW) observations of dark sirens do not directly provide the source redshift. Instead, the GW signal yields an estimate of the luminosity distance $D_L^{\rm GW}$, while the corresponding source redshift is inferred statistically using additional information, such as the redshifted source-frame mass distribution or galaxy catalogs containing potential host galaxies \cite{Abbott4, Fishbach}. This distinction is important because the redshift information associated with dark-siren GW events depends on astrophysical population assumptions or galaxy-catalog information, rather than being directly measured from the GW signal itself \cite{Abbott4, Pozzo}. In the present analysis, the GW likelihood is constructed from the GW luminosity-distance measurements, and (as shown earlier in the manuscript) our model predicts $D_L^{\rm GW}=D_L^{\rm EM}$. Consequently, combining GWTC(2.1+3+4) with CC+BAO or CC+BAO+Pantheon would mainly add another constraint on the same background distance--redshift relation already probed by the electromagnetic datasets, rather than providing an independent constraint on the tensor-propagation sector. Therefore, we analyze GWTC (2.1+3+4) separately as an independent standard-siren dataset. This allows us to assess the cosmological constraints obtained from GW luminosity distances alone and to perform a consistent comparison between the k-essence and $\Lambda$CDM models using the same GW-only data, without mixing the statistically inferred redshift information (and its associated population or galaxy-catalog assumptions) with electromagnetic redshift measurements \cite{Abbott4, Pozzo, Gray}.

The inferred parameter constraints and the corresponding 2D corner plot for the k-essence model are shown in Table~\ref{tab:II} and Fig.~\ref{Fig7}. For comparison, the analogous constraints for the standard $\Lambda$CDM model are reported in Table~\ref{tab:VIII}, with the corresponding corner plot shown in Fig.~\ref{Fig11} (see Appendix~\ref{App:E}). In particular, the GWTC(2.1+3+4)-only analysis yields essentially identical constraints on $H_0$ in both models (Tables~\ref{tab:II} and \ref{tab:VIII}). This is expected because, for a GW-only standard-siren likelihood, the dominant information comes from the background luminosity-distance relation, and in our framework this relation is effectively the same for k-essence and $\Lambda$CDM over the redshift range probed by GWTC. As a result, the inferred $H_0$ is driven primarily by the same distance-scale calibration rather than by model-specific late-time dynamics, even though the k-essence model contains additional parameters compared to $\Lambda$CDM.\\

An interesting aspect of the present framework is the link between the redshift evolution of the effective mass term $m_{\rm eff}(z)$ and the background parameters inferred from cosmological data, including the Hubble constant $H_0$. Since $m_{\rm eff}(z)$ is fixed by the k-essence background solution, the same non-canonical dynamics that sources the dispersive term in the tensor-mode equation also controls the late-time expansion history and can therefore shift the cosmological parameters preferred by background datasets relative to standard $\Lambda$CDM. In this way, the model can produce small, observationally allowed departures from $\Lambda$CDM while remaining consistent with current gravitational-wave bounds.

The evolution of $m_{\rm eff}(z)$ is a distinctive theoretical prediction of the k-essence framework. In our analysis it is not treated as an additional fit parameter; rather, it is reconstructed as a derived quantity once the primary parameters are constrained. The same background dynamics determine the expansion history and the effective dark-energy equation of state, while $m_{\rm eff}(z)$ encodes the associated dispersive modification of tensor propagation. The present results therefore provide a concrete theoretical connection between the background cosmology and the GW propagation sector.

Finally, the framework preserves luminal GW propagation, $c_{\rm gw}=1$, while still allowing dispersive phase corrections through $m_{\rm eff}$, making it phenomenologically attractive. In contrast to modified-gravity scenarios that predict $c_{\rm gw}\neq 1$ and are tightly constrained by multimessenger observations, the present model remains compatible with the GW170817 bound while permitting nontrivial late-time cosmological effects.\\

Overall, our results show that the k-essence model provides a statistically viable fit to the current observations, with predictive performance close to that of the standard $\Lambda$CDM model for the main CC+BAO+Pantheon data combination. At the same time, the k-essence framework predicts a distinctive, background-induced effective mass for tensor perturbations while preserving luminal GW propagation, $c_{\rm gw}=1$. The main significance of the model is therefore not a demonstrated resolution of the Hubble tension, but rather the possibility of linking a viable late-time expansion history to a dispersive modification of gravitational-wave propagation. Future studies may use larger standard-siren samples together with complementary early- and late-universe probes, which will be needed to test whether this characteristic GW signature can provide additional evidence for, or stronger constraints on, the k-essence scenario.

\section{Conclusion} 
\label{Sec.VII}
In this work, we have presented a combined theoretical and observational study of the propagation of linearized gravitational waves (GWs) in k-essence cosmology, in which the dark sector is described by a non-canonical scalar field. Starting from tensor perturbations on a curved FLRW background and adopting the high-frequency (short-wavelength) limit, we derived a modified gravitational-wave evolution equation that encodes the impact of the background scalar-field dynamics. Within the same homogeneous and isotropic setting, and under the weak-field approximation, we showed that this tensor-mode evolution consistently captures the influence of the non-canonical k-essence sector on GW propagation. A central outcome is the appearance of a time-dependent effective mass-like term, $m_{\rm eff}$, for tensor modes, sourced by the evolving k-essence background (and vanishing when the k-essence Lagrangian is switched off), while the background expansion enters through the standard FLRW factors in the propagation equation.

Importantly, the model preserves luminal gravitational-wave propagation ($c_{\rm gw}=1$), in agreement with stringent multimessenger bounds (e.g., GW170817), while still yielding a non-trivial modification of the dispersion relation. Specifically, the k-essence background induces an ultra-light effective mass term $m_{\rm eff}$ for tensor modes. This contribution leaves the number of propagating degrees of freedom unchanged, but generates a cumulative, frequency-dependent phase shift $\Delta\Theta$ in GW signals that builds up over cosmological distances. The resulting dispersive imprint provides a potential observational handle on the microphysics of dark energy. Moreover, we found that $m_{\rm eff}$ depends explicitly on cosmological background quantities, including the Hubble parameter and the scalar-field equation-of-state parameter $\bar{w}_{\phi}$, thereby linking GW observables directly to the underlying expansion history and k-essence dynamics.

We then specialized to a class of k-essence models with a power-law kinetic term and the quadratic potential in Eq.~\eqref{44}. Rewriting the modified gravitational-wave equation in redshift space allows a direct comparison with observations. We performed a joint Bayesian analysis using different combinations of background-expansion probes, Cosmic Chronometers (CC), BAO, and Pantheon+ supernovae, and, where applicable, GW standard-siren information from the GWTC-2.1, GWTC-3, and GWTC-4 catalogs. Overall, the model is compatible with current constraints and reproduces the observed late-time expansion history. The CC+BAO+Pantheon+ combination provides the tightest constraints on the present-day parameters and yields our best-fit values. For the derived effective mass, we obtain $m_{\rm eff,0}\equiv m_{\rm eff}(z=0)\approx 212.88\pm 1.72\, \mathrm{km\,s^{-1}\,Mpc^{-1}}$, which corresponds to an ultra-light physical scale of $\mathcal{O}(10^{-33})\,\mathrm{eV}$ set by the underlying k-essence background.

To quantify statistical performance, we compared the k-essence model with a $\Lambda$CDM control model using standard goodness-of-fit and model-selection diagnostics. The resulting $\chi^2$ and reduced $\chi^2$, posterior-predictive checks, and information criteria (AIC, BIC, DIC) indicate that the k-essence framework is statistically viable and fits the combined CC+BAO+Pantheon+ dataset at a level comparable to $\Lambda$CDM. While the current background data do not require the additional k-essence degrees of freedom, they remain consistent with the observations and offer an alternative theoretical framework for gravitational-wave propagation.

Overall, this study highlights gravitational waves as precision probes of the cosmological background and shows that dispersive propagation provides a robust, testable signature of non-canonical scalar-field (k-essence) models. The framework developed here offers a consistent and observationally viable extension of standard cosmology, in which departures from general relativity manifest mainly as phase corrections rather than as changes in wave speed. In this way, fine dispersive features in gravitational-wave signals can be used to probe the microphysics of dark energy while remaining fully compatible with a luminal propagation speed.

Several directions could extend and deepen this analysis. First, going beyond linear perturbation theory to second order would allow a systematic study of scalar-tensor couplings. Second, exploring more general k-essence scenarios, or embedding the framework within Horndeski and beyond-Horndeski (DHOST) theories, could reveal richer phenomenology, including possible modifications to GW amplitudes and polarizations and a non-trivial redshift evolution of the effective mass induced by non-minimal couplings.

Third, a waveform-level treatment within full parameter-estimation pipelines for current ground-based detectors (LIGO, Virgo, and KAGRA), as well as future observatories (LISA, Taiji, TianQin, the Einstein Telescope, and Cosmic Explorer), will be essential to assess the detectability of the predicted cumulative phase shifts and to constrain key parameters such as the kinetic index $n$ and the potential $V(\phi)$. Finally, combining GW measurements with large-scale-structure probes and cosmic microwave background (CMB) data may help break parameter degeneracies. Extending the framework to early-universe applications, including inflationary tensor modes and stochastic GW backgrounds, could further illuminate the origin and evolution of the dark sector. These directions are left for future work and are beyond the scope of the present analysis.\\

\begin{acknowledgments}
We thank the anonymous referee for constructive comments and suggestions, which have substantially improved the manuscript. G.M. (IUCAA Associate) and S.B. acknowledge the Inter-University Centre for Astronomy and Astrophysics (IUCAA), Pune, India, for hosting them and supporting part of this work during their visit. G.M. and E.G. thank the COST Association (CA21136 CosmoVerse), European Union, for the opportunity to participate in this international association as group members. E.G. also thanks the COST Association (CA23130, BridgeQG), European Union, for the opportunity to participate in this international association as a group member. R.H. acknowledges the financial support of the  Agencia Nacional de Investigación y Desarrollo, ANID / Concurso FOVI-250062, 2025. A.B. acknowledges partial support from the RUSA project at the University of Calcutta. G.M. extends his gratitude to all undergraduate, postgraduate, and doctoral students, along with his teachers, collaborators, well-wishers, and his son (Aditya), whose support has significantly enriched his academic journey. S.B. gratefully acknowledges his parents for their constant love, encouragement, and unwavering support throughout every stage of his academic and personal journey.\\

\textbf{Conflicts of interest:} The authors declare no conflicts of interest.\\

\textbf{Funding information:} Not available.\\

\textbf{Data availability:} The data used in this study are readily accessible from public sources for validation of our model; however, we did not generate any new data sets for this research.\\

%\textbf{Declaration of competing interests:} The authors declare that they have no known competing financial interests or personal relationships that could have appeared to influence the work reported in this paper.\\

%\textbf{Declaration of generative AI in scientific writing:} \textcolor{blue}{Overleaf AI assistance was used exclusively for language editing and improving readability. It was not used for scientific analysis, derivations, interpretation, or data processing. The authors take full responsibility for the scientific content of the manuscript.}
\end{acknowledgments}

\appendix
\section{Derivations of k-essence Theory} \label{App:A}
The k-essence action is
\ben
S_k[\phi, g^{\mu\nu}]=\int d^4x \sqrt{-g}\La(X, \phi) \label{A1}
\een
where the canonical kinetic term $X=\frac{1}{2}g^{\mu\nu}\nabla_\mu\phi \nabla_\nu\phi$. As $\phi$ is a scalar filed, so er can write, $X=\frac{1}{2}g^{\mu\nu}\partial_\mu\phi \partial_\nu\phi$.\\

Now the variation of the above action $S_k[\phi, g^{\mu\nu}]$ can be derived in two parts, one with respect to with respect to the metric $g^{\mu\nu}$ and another with respect to $\phi$  as :
\ben
\delta S_k[\phi, g^{\mu\nu}]=\delta S_k[\phi, g^{\mu\nu}]|_{g^{\mu\nu}}+\delta S_k[\phi, g^{\mu\nu}]|_\phi \label{A2}
\een
Here $\delta S_k[\phi, g^{\mu\nu}]|_{g^{\mu\nu}}$ and $\delta S_k[\phi, g^{\mu\nu}]|_\phi$ are the variations of the k-essence action $S_k[\phi, g^{\mu\nu}]$ with respect to the metric $g^{\mu\nu}$ and the scalar field $\phi$ respectively. 

Now the first term in \ref{A2} can be expressed as, 
\ben
\delta S_k[\phi, g^{\mu\nu}]|_{g^{\mu\nu}}=\int d^4x \sqrt{-g} \delta g^{\mu\nu} \frac{1}{2}\Big[\La_X \nabla_\mu \phi \nabla_\nu \phi-g^{\mu\nu}\La\Big] \nonumber\\ \label{A3}
\een

Therefore, the energy-momentum tensor $T_{\mu\nu}$ is:
\ben
&&T_{\mu\nu}=-\frac{2}{\sqrt{-g}}\frac{\delta S_k[\phi, g^{\mu\nu}]|_{g^{\mu\nu}}}{\delta g^{\mu\nu}}\nonumber\\
\text{or,}&&T_{\mu\nu}= g_{\mu\nu}\La-\La_X \nabla_\mu \phi \nabla_\nu \phi \label{A4}
\een
As $\phi$ is a scalar field then we also can write \cite{gm2}
\ben
T_{\mu\nu}= g_{\mu\nu}\La-\La_X \partial_\mu \phi \partial_\nu \phi \label{A5}
\een

On the other hand, the second term of \ref{A2} can be expressed as:
\ben
\delta S_k[\phi, g^{\mu\nu}]|_\phi&=&-\int d^4x \sqrt{-g} \Big[(\La_{XX} \nabla^\mu \phi \nabla^\nu \phi \nonumber\\ &+& \La_X g^{\mu\nu})\nabla_\mu\nabla_\nu\phi+2X\La_{X\phi}-\La_\phi \Big]\delta\phi \nonumber\\ \label{A6}
\een
where $\La_X\equiv \frac{\partial\La}{\partial X}$, $\La_{X\phi}\equiv \frac{\partial^2\La}{\partial X \partial \phi}$ , $\La_{XX}\equiv \frac{\partial^2\La}{\partial^2 X}$ and $\La_{\phi}\equiv \frac{\partial\La}{\partial \phi}$.

Therefore, the scalar field equation of motion is \cite{gm2}:
\ben
-\frac{1}{\sqrt{-g}}\frac{\delta S_k}{\delta \phi}=\bar{G}^{\mu\nu} \nabla_\mu \phi \nabla_\nu \phi+ 2X \La_{X\phi}-\La_\phi \label{A7}
\een
where $\bar{G}^{\mu\nu}$ is the effective emergent metric whis can be defined as:
\ben
\bar{G}^{\mu\nu}=g^{\mu\nu}\La_X+\La_{XX}\nabla^\mu \phi \nabla^\nu \phi \label{A8}
\een
with $1+\frac{2X\La_{XX}}{\La_X}>0$ and the velocity of sound $c^2_s=\Big(1+2X\frac{\La_{XX}}{\La_X}\Big)^{-1}$ \cite{gm2}. Here ``Bar" $(\bar{})$ sign does not indicate the background geometry.

Now the inverse of the effective emergent metric is expressed as:
\ben
\bar{G}_{\mu\nu}=\frac{\La_X}{c_s} \Big[g_{\mu\nu}-c^2_s\frac{\La_{XX}}{\La_X}\nabla_\mu \phi \nabla_\nu \phi \Big]\label{A8}
\een
Using the relation $\tilde{G}_{\mu\nu}=\frac{c_s}{\La_X}\bar{G}_{\mu\nu}$, we get
\ben
\tilde{G}_{\mu\nu}=g_{\mu\nu}-\frac{\La_{XX}}{\La_X+2X\La_{XX}}\nabla_\mu \phi \nabla_\nu \phi
\een
Here ``Tilde"$(\tilde{})$ sign is not indicating the trace reversed terms.

\section{Non-Zero Curvature tensors From Background Universe } \label{App:B}
In section\ref{subsec:B} we have already considered the conformal flat FLRW universe with scalar-tensor perturbation in Eq.(\ref{28}). The background spacetime metric is (already mentioned in Eq.(\ref{30}))   
\ben
d\bar{s}^2= \bar{g}_{\mu\nu} dx^\mu dx^\nu= a^2(\eta)[d\eta^2-\delta_{ij} dx^i dx^j] \label{B1}
\een

Here $a^2(\eta)$ is the scale factor, and the conformal time $\eta$ is defined in form of cosmic time coordinate $t$ as follows, \cite{Garriga, Dodelson}
\ben
\eta \equiv \int \frac{dt}{a(t)} \label{B2}
\een
The non-zero background connection coefficients are
\ben
&\bar{\Gamma}^0_{00}=&(a'/a)=\mathcal{H}\nonumber\\
&\bar{\Gamma}^0_{ij}=&\mathcal{H}\delta_{ij}\nonumber\\
&\bar{\Gamma}^i_{0j}=&\mathcal{H}\delta^i_j \label{B3}
\een
So the non-zero components of background Riemann tensors can be calculated as:
\ben
\bar{R}_{0i0j}&=&\Big(aa''-a'^2\Big) \delta_{ij}=a^2\mathcal{H}'\delta_{ij}\nonumber\\
\bar{R}_{i00j}&=&-a^2 \mathcal{H}'\delta_{ij}\nonumber\\
\bar{R}_{ijkl}&=&-a'^2(\delta_{ik}\delta_{jl}-\delta_{il}\delta_{jk})\nonumber\\&=&-a^2\mathcal{H}^2(\delta_{ik}\delta_{jl}-\delta_{il}\delta_{jk}) \label{B4}
\een
The non-zero components of the background Ricci tensors are
\ben
&\bar{R}_{00}&=-3\Big(\frac{a''}{a}-\frac{a'^2}{a^2}\Big)=-3\mathcal{H}'\nonumber\\
&\bar{R}_{ij}&=\Big(\frac{a''}{a}+\frac{a'^2}{a^2}\Big)\delta_{ij}=(\mathcal{H}'+2\mathcal{H}^2)\delta_{ij} \label{B5}
\een
and the background Ricci scalar is
\ben
\bar{R}=-\frac{6}{a^2}\Big(\frac{a''}{a}\Big)=-\frac{6}{a^2}(\mathcal{H}'+\mathcal{H}^2) \label{B6}
\een

Here $\mathcal{H}$ is the conformal or comoving Hubble parameter, i.e. $\mathcal{H}\equiv\mathcal{H}(\eta)$. It can be defined in form of $a(\eta)$ as follows:
\ben
\mathcal{H} \equiv \Big(\frac{a'}{a}\Big) \label{B7}
\een
and note that,
\ben
\mathcal{H}' \equiv \Big(\frac{a'}{a}\Big)'= \frac{a''}{a}-\Big(\frac{a'}{a}\Big)^2= \frac{a''}{a}-\mathcal{H}^2 \label{B8}
\een
where ``prime" $(' \equiv \partial_\eta)$ denotes the first order derivative with respect to the conformal time.\\

Now here we consider an additional matter action $S_m$ with the following definition, $S_m=\int d^4x \sqrt{-g}\La_m$, where $\La_m$ is a matter Lagrangian. Therefore the total action will be,
\ben
S_{total}=S+S_m \label{B9}
\een
where $S$ is the action that is already mention in Eq.(\ref{16}).

Now according to Ref.\cite{Padmanabhan}, we have the following total background energy-momentum tensor for perfect fluid consideration,
\ben
\bar{T}_{\mu\nu}=(\bar{\rho}+\bar{P})\bar{u}_{\mu}\bar{u}_{\nu}-\bar{P}\bar{g}_{\mu\nu} \label{B10}
\een
where the background four-velocity vector $\bar{u}_\mu=\frac{\partial_\mu \phi_0}{\sqrt{2\bar{X}}}$, which satisfy the normalization condition, $\bar{u}_\mu\bar{u}^\nu=1$, $\bar{\rho}=(\bar{\rho}_\phi+\bar{\rho}_m)$ and $\bar{P}=(\bar{P}_\phi+\bar{P}_m)$ is the background energy density and pressure respectively. Here $\bar{\rho}_\phi$ and $\bar{P}_\phi$ is the k-essence background scalar field energy density and pressure (already defined in Eq.(\ref{46})) respectively.

So using the conformal flat FLRW background metric in Eq.(\ref{B1}), and the pressureless dark matter, according to the cold dark matter (CDM) model, i.e. $\bar{P}_m=0$,  we express the following form of the conformal Friedmann equations,
\ben
&\mathcal{H}^2=&\frac{8\pi G}{3}a^2\bar{\rho}= \frac{8\pi G}{3}a^2(\bar{\rho}_\phi+\bar{\rho}_m) \nonumber\\
\text{and} &\mathcal{H}'=&-\frac{4\pi G}{3}a^2(\bar{\rho}+3\bar{P})=-\frac{4\pi G}{3}a^2(\bar{\rho}_\phi+\bar{\rho}_m+3\bar{P}_\phi)\nonumber\\ \label{B11}
\een

\section{Dimensional analysis of the effective mass term $m_{\rm eff}$}\label{App:C}
To verify the physical consistency of the modified gravitational-wave equation~\eqref{40}, we use dimensional analysis to confirm that the background-induced quantity $m_{\rm eff}^2$ defined in Eq.~\eqref{48} carries dimensions of (mass)$^2$. Working in natural units $c=\hbar=1$, one has
\ben
[L]=[T]=[M]^{-1}=[E]^{-1}
\label{c1}
\een
in the relativistic limit \cite{Zee, Peskin, Schwartz, Weinberg}.

The total background-induced potential term introduced in Eq.~\eqref{47a} (equivalently Eq.~\eqref{47b}) is
\ben
U^2(\eta)= 8\mathcal{H}'+10\mathcal{H}^2+16\pi G a^2\,\La(\bar{X},\bar{\phi})\,.
\label{c2}
\een
Here $a$ is dimensionless, a prime denotes $d/d\eta$, and any derivative operator (e.g. $\partial_\mu\equiv\partial/\partial x^\mu$) has mass dimension $+1$.

Since $\mathcal{H}=a'/a$ has dimensions of inverse conformal time, $[\mathcal{H}]\sim[T]^{-1}\sim[M]$, which implies
\ben
[\mathcal{H}^2]\sim[M]^2,\qquad [\mathcal{H}']\sim[M]^2\,.
\label{c3}
\een
Moreover, Newton's constant satisfies $[G]\sim[M]^{-2}$ in natural units, while the Lagrangian density has the dimensions of an energy density, $[\La]\sim[E]^4\sim[M]^4$ in four spacetime dimensions \cite{Zee, Peskin, Schwartz, Weinberg}. Therefore,
\ben
[16\pi G a^2\La]\sim [G]\,[a^2]\,[\La]\sim [M]^{-2}\times 1\times [M]^4\sim[M]^2\,.
\label{c4}
\een
All terms contributing to $U^2$ in Eq.~\eqref{c2} thus scale as $[M]^2$. Since $m_{\rm eff}^2$ in Eq.~\eqref{48} is obtained from $U^2$ by subtracting the matter contribution $16\pi G a^2\bar\rho_m$, which also has dimension $[M]^2$, it follows that $m_{\rm eff}^2$ indeed behaves dimensionally as an effective mass-squared parameter.

\section{Observational Data Sets} \label{App:D}
The Hubble parameter values $H(z)$  listed in Table (\ref{tab:II}), are collected from various observational studies covering a redshift range approximately from $z\sim0.07$ to $z\sim1.965$. These measurements are obtained using the cosmic chronometer (or differential age) technique, which determines $H(z)$ by evaluating the age difference between passively evolving galaxies at distinct redshifts.

\begin{table}[H]
    \centering
    \begin{tabular}{|c|c|c|c|c|}
    \hline
       $z$ & \textbf{$H(z)$} & \textbf{$\sigma_{H,i}$}  & \textbf{Refs.} \\
        \hline
        0.07 & 69.0 & $\pm 19.60$  &  \cite{zhang}\\
        \hline
         0.0708 & 69.0 & $\pm 19.68$  & \cite{Abbott1}\\
        \hline
         0.09 & 69.0 & $\pm 12.0$  & \cite{Jimenez1}\\
        \hline
         0.12 & 68.6 & $\pm 26.2$  &  \cite{zhang}\\
        \hline
         0.17 & 83.0 & $\pm 8.00$ &  \cite{Simon}\\
        \hline
         0.179 & 75.0 & $\pm 4.0$  & \cite{Moresco}\\
        \hline
         0.199 & 75.0 & $\pm 5.0$  & \cite{Moresco}\\
        \hline
         0.20 & 72.9 & $\pm 29.6$  & \cite{Abbott1}\\
        \hline
        
         0.27 & 77.0 & $\pm 14.0$  &  \cite{Simon}\\
        \hline
         0.28 & 88.8 & $\pm 36.6$  & \cite{Abbott1}\\
        \hline
        
         0.352 & 83.0 & $\pm 14.0$  &  \cite{Moresco}\\
        \hline
         0.38 & 81.2 & $\pm 2.2$  & \cite{Metin}\\
        \hline
         0.3802 & 83.0 & $\pm 14.0$  & \cite{Eric}\\
        \hline
         0.4 & 95.0 & $\pm 17.0$  &  \cite{Simon}\\
        \hline
         0.4004 & 77.0 & $\pm 10.2$  & \cite{Eric}\\
        \hline
         0.4247 & 87.1 & $\pm 11.2$ & \cite{Eric}\\
        \hline
         
         0.4497 & 92.8 & $\pm 12.9$  & \cite{Eric}\\
        \hline
         0.47 & 89.0 & $\pm 50.0$ & \cite{Julian}\\
        \hline
         0.4783 & 80.9 & $\pm 9.0$  &  \cite{Eric}\\
        \hline
         0.48 & 97.0 & $\pm 62.0$  &  \cite{Stern}\\
        \hline
         0.51 & 90.9 & $\pm 2.1$  & \cite{Metin}\\
        \hline
         
         0.593 & 104.0 & $\pm 13.0$  & \cite{Moresco}\\
        \hline
        
         0.61 & 98.96 & $\pm 2.20$  & \cite{Metin}\\
        \hline
         0.68 & 92.00 & $\pm 8.0$  &  \cite{Moresco}\\
        \hline
         
         0.781 & 105.0 & $\pm 12.0$  &  \cite{Moresco}\\
        \hline
         0.875 & 125.0 & $\pm 17.0$  &  \cite{Moresco}\\
        \hline
         0.88 & 99.0 & $\pm 40.0$  &  \cite{Stern}\\
        \hline
         0.9 & 117.0 & $\pm 23.0$  &  \cite{Simon}\\
        \hline
         1.037 & 154.0 & $\pm 20.0$  & \cite{Moresco}\\
        \hline
         1.3 & 168.0 & $\pm 17.0$  & \cite{Simon}\\
        \hline
         1.363 & 160.0 & $\pm 33.6$  & \cite{Betoule}\\
        \hline
         1.43 & 177.0 & $\pm 18.0$  &  \cite{Simon}\\
        \hline
         1.53 & 140.0 & $\pm 14.0$  &  \cite{Simon}\\
        \hline
         1.75 & 202.0 & $\pm 40.0$  &  \cite{Simon}\\
        \hline
         1.965 & 186.5 & $\pm 50.4$  & \cite{Betoule}\\
        \hline
    \end{tabular}
    \caption{Here the $H(z)$ value is derived from the cosmic chronological method or the differential age approach, with the relevant references cited accordingly. The unit of $H(z)$ is $km.s^{-1}.Mpc^{-1}$, 'CC' indicates that .}
    \label{tab:V}
\end{table}

We analyze two separate BAO datasets in this study. The first dataset, referred to as SDSSBAO, consists of 8 data points obtained from the SDSS. The second dataset, called DESIBAO, includes 9 data points from the DESI. For the DESI dataset, we specifically use the most recent DR2 provided by the collaboration. The combination of these two datasets, SDSSBAO and DESIBAO, is collectively referred to as the BAO dataset in our analysis. The detailed values for both datasets are presented in Table (\ref{tab:III}) and Table (\ref{tab:IV}).

\renewcommand{\arraystretch}{1.2}
\begin{table*}[t]
    \centering
    \begin{tabular}{|c|c|c|c|c|c|c|c|c|}
    \hline
        \textbf{$\Tilde{z}$} & \textbf{0.15} & \textbf{0.38} & \textbf{0.51} & \textbf{0.70} & \textbf{0.85} & \textbf{1.48} & \textbf{2.33} & \textbf{2.33} \\
        \hline
       $D_V(\Tilde{z})/r_d$ & $4.47\pm 0.17$ & - & - & -&$18.33^{+0.57}_{-0.62}$ & - & - & - \\
        \hline
      $D_M(\Tilde{z})/r_d$ & - & $10.23\pm 0.17$ & $13.36\pm 0.21$ & $17.86\pm 0.33$ & - & $30.69\pm 0.80$ & $37.6\pm 1.9$ & $37.3\pm 1.7$ \\
        \hline
        $D_H(\Tilde{z})/r_d$ & - & $25.00\pm 0.76$ & $22.33\pm 0.58$ & $19.33\pm 0.53$ & - & $13.26\pm 0.55$ & $8.93\pm 0.28$ & $9.08\pm 0.34$ \\
        \hline
    \end{tabular}
    \caption{SDSSBAO Dataset}
    \label{tab:VI}
\end{table*} 

\renewcommand{\arraystretch}{1.2}
\begin{table}[t]
    \centering
    \begin{tabular}{|c|c|c|c|}
    \hline
        \textbf{$\Tilde{z}$} & \textbf{ $D_V(\Tilde{z})/r_d$} & \textbf{ $D_M(\Tilde{z})/r_d$} & \textbf{$D_H(\Tilde{z})/r_d$} \\
        \hline
        $0.295$ & $7.942\pm 0.075$ &- &- \\
        \hline
        $0.510$ & $12.720\pm 0.099$ & $13.588\pm 0.167$ & $21.863\pm 0.425$ \\
        \hline
        $0.706$ & $16.050\pm 0.110$ & $17.351\pm 0.177$ & $19.455\pm 0.330$ \\
        \hline
        $0.922$ & $19.656\pm 0.105$ & $21.648\pm 0.178$ & $17.577\pm 0.213$ \\
        \hline
        $0.934$ & $19.721\pm 0.091$ & $21.576\pm 0.152$ & $17.641\pm 0.193$ \\
        \hline
        $0.955$ & $20.008\pm 0.183$ & $21.707\pm 0.335$ & $17.803\pm 0.297$ \\
        \hline
        $1.321$ & $24.252\pm 0.174$ & $27.601\pm 0.318$ & $14.176\pm 0.221$ \\
        \hline
        $1.484$ & $26.055\pm 0.398$ & $30.512\pm 0.760$ & $12.817\pm 0.516$ \\
        \hline
        $2.330$ & $31.267\pm 0.256$ & $38.988\pm 0.531$ & $8.632\pm 0.101$ \\
        \hline
    \end{tabular}
    \caption{DESI BAO Dataset (DR2)}
    \label{tab:VII}
\end{table}

\begin{figure}[H]
\centering
\includegraphics[width=7.8cm, height=6.5cm]{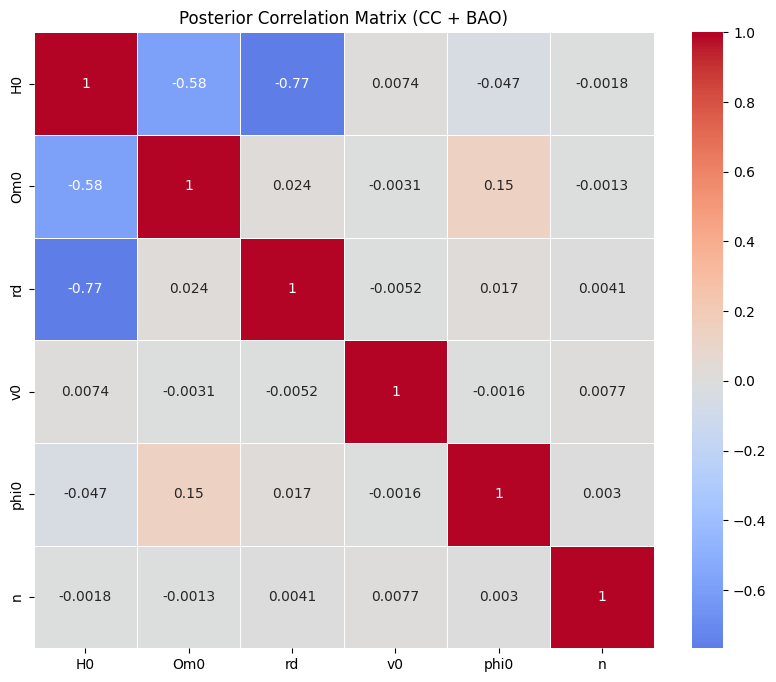}
 \caption{Bayesian inference correlation heatmap structure for the observational datasets CC + BAO in k-essence model.}\label{Fig8} 
\end{figure}

\begin{figure}[H]
\centering
\includegraphics[width=7.8cm, height=6.5cm]{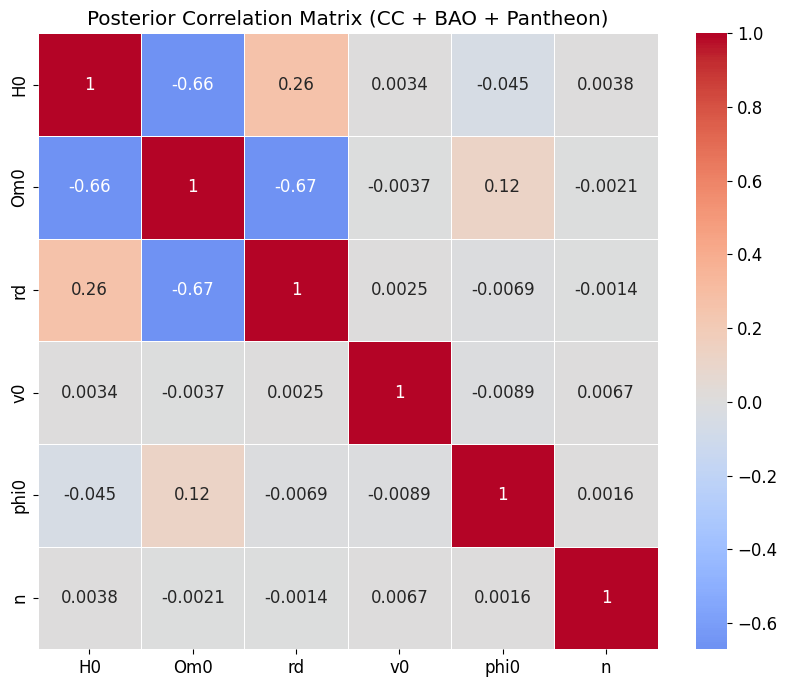}
 \caption{Bayesian inference correlation heatmap structure for the observational datasets CC + Pantheon + BAO in k-essence model.}\label{Fig9} 
\end{figure}

\section{$\Lambda$CDM model} \label{App:E}
We consider the standard $\Lambda$CDM model and perform a Bayesian analysis to compare it with our proposed k-essence model using various combinations of CC, BAO, and Pantheon+ datasets. In the $\Lambda$CDM model, the energy content of the Universe is assumed to be dominated by two components: pressureless cold dark matter with energy density $\rho_m$ and dark energy with energy density $\rho_{\rm de}$.

Assuming a spatially flat FLRW universe, the governing differential equations for the Hubble parameter $H(z)$ and the matter density parameter $\Omega_m(z)$ are

\ben
\frac{dH}{dz}=\frac{3H(z)}{2(1+z)}\Omega_m(z)\label{E1}
\een

\ben
\frac{d\Omega_m}{dz}=-\frac{2\Omega_m}{H}\Big(\frac{dH}{dz}\Big) + \frac{3\Omega_m}{(1+z)}\label{E2}
\een

Here we have imposed the spatial flatness condition $\Omega_{m,0}+\Omega_{\rm de,0}=1$.

\begin{figure}[H]
\centering
\includegraphics[width=7.9cm, height=7.8cm]{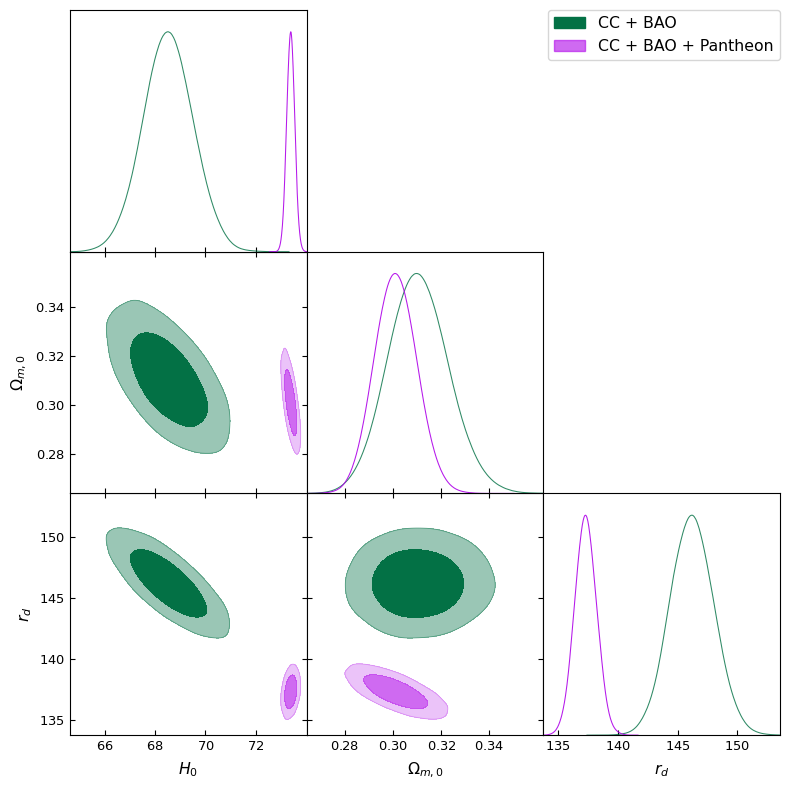}
 \caption{Fitting of cosmological parameters in the $\Lambda$CDM model using various combinations of CC, Pantheon, and BAO datasets.
}\label{Fig10} 
\end{figure}

\begin{table*}[t]
    \centering
    \renewcommand{\arraystretch}{1.3}
    \caption{Cosmological parameter constraints in the $\Lambda$CDM model for different datasets.}
     \label{tab:VIII}
     \begin{tabular}{c c c c c }
        \hline\hline
        \textbf{Parameters} & \textbf{Prior} & \multicolumn{3}{c|}{\textbf{Datasets}} \\ \cline{3-5} 
 & & \textbf{CC + BAO } & \textbf{CC + Pantheon + BAO } & \textbf{GWTC (2.1+3+4) only}\\ \hline
$H_0$ [Km/s/Mpc] & $\mathcal{U}(60, 80)$ & $68.5 \pm 1.0$ & $73.39 \pm 0.16$ & $68.1 \pm 1.7$\\ 
$\Omega_{m,0}$ & $\mathcal{U}(0.2, 0.4)$ & $0.310 \pm 0.013$ & $0.301 \pm 0.0088$ & $0.303 \pm 0.056$ \\ 
$r_d$ [Mpc] & $\mathcal{U}(130, 160)$ & $146.2 \pm 1.8$ & $137.28 \pm 0.91$ & --\\
    \hline\hline
    \end{tabular}
\end{table*}

\begin{figure}[H]
\centering
\includegraphics[width=7.9cm, height=7.8cm]{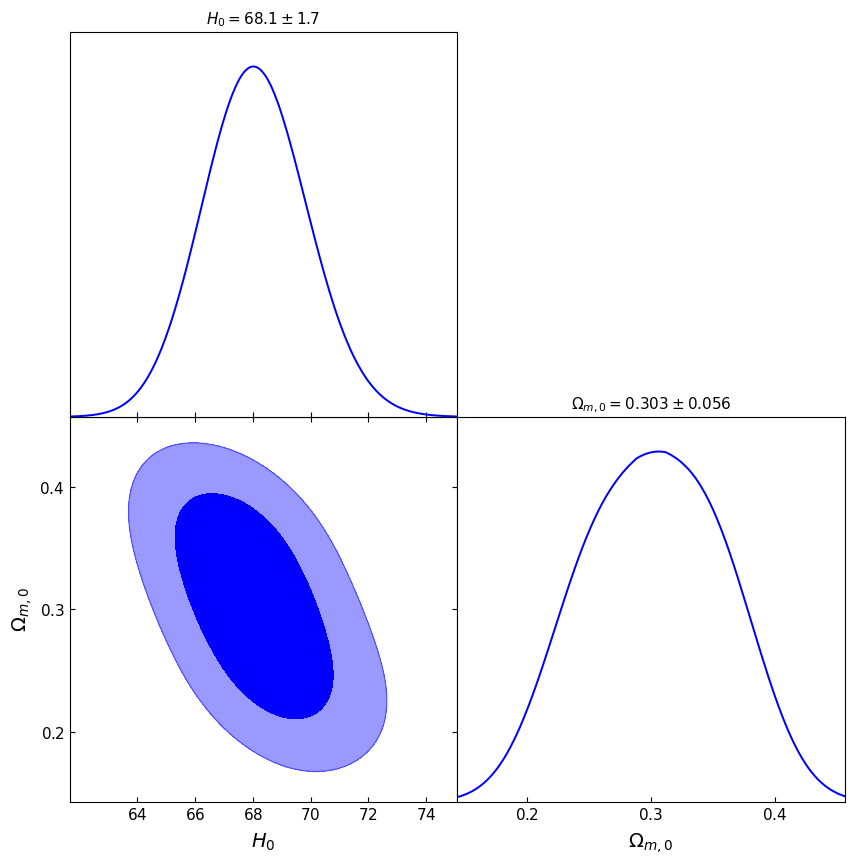}
 \caption{Fitting of cosmological parameters in $\Lambda$CDM model with GWTC data (GWTC 2.1+3+4)}\label{Fig11} 
\end{figure}

\end{document}